\documentclass[%
 reprint,
 amsmath,amssymb,
 aps,
pre,
floatfix,
]{revtex4-2}

\usepackage{graphicx}
\usepackage{dcolumn}
\usepackage{bm}
\usepackage{hyperref}
\usepackage{enumitem}
\usepackage{float}

\DeclareMathOperator{\Tr}{Tr}

\begin{document}

\preprint{APS/123-QED}

\title{A hyperelastic theory for nonlinear hydrogel diffusiophoresis}

\author{Chinmay Katke$^{1, 2}$}
\author{C. Nadir Kaplan$^{1-4}$}
 \email{nadirkaplan@vt.edu}
\affiliation{%
 $^1$Department of Physics,
Virginia Polytechnic Institute and State University, Blacksburg, VA 24061, USA. 
\\$^2$Center for Soft Matter and Biological Physics,
Virginia Polytechnic Institute and State University, Blacksburg, VA 24061, USA. 
\\$^3$Department of Mechanical Engineering,
Virginia Polytechnic Institute and State University, Blacksburg, VA 24061, USA. 
\\$^4$Center for the Mathematics of Biosystems,
Virginia Polytechnic Institute and State University, Blacksburg, VA 24061, USA.
}

\date{\today}

\begin{abstract}
Hydrogel diffusiophoresis is the deformation of a hydrogel due to a solute gradient that leads to a gradient of pairwise interactions between the solute particles and the hydrogel polymers to trigger osmotic flux. Unlike typical osmosis, it occurs without any interface selectivity of the gel to the solute and can overcome the diffusive swelling without any structural modifications to the gel. We have recently shown this effect for linear deformations of a chemically responsive polyacrylic acid (PAA) hydrogel that releases ions upon arrival of a stimulus (acid), thus internally generating the solute gradient required for diffusiophoresis [Phys. Rev. Lett. 132, 208201 (2024)]. Here we develop a nonlinear poroelastic theory for large diffusiophoretic gel strains in two models: Model I considers deformations of a generic gel when an external solute gradient is imposed and indicates that deformations can be stored while the stimulus gradient persists. Model II is concerned with the internal generation of the solute gradient within the gel, motivated by the coupled PAA gel, solute (copper), and stimulus (acid) system. In Model II, we investigate the nonlinear deformations for high stimulus concentrations or by changing the solute particle size to boost steric polymer-solute interactions, as well as under a stimulus flow through the gel. Compared to the experimental strain rates in Katke [Phys. Rev. Lett. 132, 208201 (2024)], Model II demonstrates that varying the stimulus concentration can increase the strain rate up to four times, changing the solute particle size up to $\sim 25$ times, and imposed flow up to $\sim 40$ times. Our theory couples nonlinear poroelasticity, polymer-solute interactions, and reaction-transport dynamics to predict large and fast diffusiophoretic gel deformations, which may find applications in hydrogel-based soft robotics and drug delivery.
\end{abstract}

\maketitle

\section{\label{sec:intro} Introduction}
Hydrogels can swell and retain water due to the inherent osmotic pressure of a constituent crosslinked polymer network, enabling them to undergo large deformations in response to environmental stimuli \cite{AHMED2015105, Liu_2022, Jeon2017, Peng2018, banerjee2018}. The magnitude of deformations together with the responsiveness to different external fields can be fine-tuned by a wide range of polymer compositions available to synthesize gels. Furthermore, their structural similarity with biological tissues has resulted in applications across diverse fields such as drug delivery, tissue engineering, and biosensing \cite{YOSHIDA2018361, HOARE20081993, Ghasemiyeh2019, KESHARWANI2021102914, KHAN2024127882, Luo2024, Thai2024, JIANGLUO2024215874, QIU2001321}. In particular, the development of hydrogel-based actuators for soft robotics can be tailored to react to a variety of stimuli, including thermal, chemical, optical, electrical, hydraulic, and pneumatic triggers, thus mimicking the deformations of biological tissues~\cite{LEE2020100258, Jiao2022, Dong2006, gao2017multi, zheng2015tough, duan2017bilayer, li2017, Xu2022, li2017fast, mitchell2019easy, Yuk2017, Sun2023}.

One limiting factor for the rate of hydrogel deformations used for actuation is that the change of solvent concentration in a gel due to absorption or desorption typically obeys diffusive dynamics. For a gel with a shortest dimension $H$, an emergent diffusive time scale for gel deformations is thus given as $\tau \equiv H^2\mu_f/k_f E\,,$ where $\mu_f$ is the kinematic viscosity of the solvent, $E$ and $k_f$ are the Young's modulus and the hydraulic permeability of the hydrogel, respectively. Therefore, while micron-sized gels can deform within seconds in response to a stimulus, scaling up these designs to visible lengths while preserving fast response presents a significant challenge \cite{duan2017bilayer, Zhang2017, li2017}. Although diffusive scaling suggests an increase in the deformation rate with bigger pore size (higher $k_f$), the polymer density is reduced by $\sim k_f^{3/2}\,,$ compromising on gel functionalization and responsiveness to external fields~\cite{Choudhary2022,Arens2017}. This especially restricts the design space of chemically responsive hydrogels, which convert released energy from the exothermic reactions or intermolecular interactions of the constituent polymers with solute species into mechanical work~\cite{epstein2014,Shim2012, selfwalkinggel, duan2017bilayer}.
    
An exception to the common diffusive behavior is a PAA hydrogel system that deforms at a rate faster than $\tau^{-1}$ via an interplay between two competing chemical stimuli~\cite{Katke_2024,Korevaar2020}. PAA gel can store divalent copper or calcium ions (the first stimulus) by forming, e.g., a stable COO$^-$--Cu$^{2+}$--COO$^-$ complex~ \cite{Palleau2013}. Upon adding acid (e.g., HCl) as a second stimulus, dissociated protons replace the divalent ions by forming COOH groups. Importantly, right after the release of the first stimulus, a transient swelling spike is observed, which was found to be due to a temporary osmotic influx generated by the ion release although the gel interface is not selective to the ions or acid~\cite{Korevaar2020}. The gel returns to its original height after the ion concentration equilibrates across the gel and the supernatant domain. Control experiments have shown that, first, adding copper to the HCl solution suppresses swelling by reducing the gel hypotonicity, second, the gel height does not change when the same amount of acid is delivered over a much longer time, leading to a gradual release of the ions~\cite{Korevaar2020}. These results suggest that swelling is a dynamic response related to the magnitude of the temporary free ion gradient.

To explain the dynamics of the swelling spike, we proposed the ``gel diffusiophoresis" mechanism (Fig.~\ref{fig:diffusiophoresis})~\cite{Katke_2024}: Interactions between the free ions inside the gel fluid and polymers drive a diffusiophoretic motion of the polymer network due to the ion concentration gradient akin to the diffusiophoresis of colloidal particles~\cite{Derjaguin1947, Derjaguin1987, DERJAGUIN1993138, Marbach2019}. For the PAA gel, copper (or calcium), and acid system, we assumed that diffusiophoresis is induced by the steric repulsion between the ions and polymers with an exclusion radius $R_e.$ We found that gel diffusiophoresis can drive superdiffusive motion of the polymer network with the corresponding deformation rate for repulsive interactions given as $\tau_{DP}^{-1}\equiv\frac{k_BT R_e^2}{2 H^2\mu_f v_c} \gg \tau^{-1}$ ($k_B:$ Boltzmann's constant, $T:$ temperature, $v_c:$ molecular volume), with the potential of enabling faster actuations.

While in Ref.~\cite{Katke_2024} we developed a linear elastic framework for small strains observed in experiments, often synthetic gels or their biological counterparts experience large, nonlinear strains~\cite{Wessendorf_2012, roan2011we, Sun2012}. Thus, herein we extend the theory of gel diffusiophoresis to nonlinear deformations by constructing two models: In Model I, a generic gel undergoes diffusiophoretic deformations in response to an externally maintained solute gradient (Fig.~\ref{fig:ch2_acid_throughflow_schematic}a). Model II is based on the PAA hydrogel, solute (copper), and stimulus (acid) system, where diffusiophoresis emerges from an internally generated solute gradient upon arrival of the stimulus (Fig.~\ref{fig:ch2_acid_throughflow_schematic}b-f). In Model II, we investigate three distinct scenarios: Scenarios 1 and 2 concern the emergence of nonlinear deformations for high stimulus concentrations and by changing the size of the solute particles to boost steric polymer-solute interactions, respectively  (Fig.~\ref{fig:ch2_acid_throughflow_schematic}b-d). Scenario 3 considers stimulus flow that is driven by a constant pressure drop imposed across the gel and the supernatant domain to generate elevated strain rates (Fig.~\ref{fig:ch2_acid_throughflow_schematic}e,~f).
    
This paper is organized as follows: In Section~\ref{sec:diffusiophoresis}, we describe the gel diffusiophoresis mechanism. In Section~\ref{sec:methods}, we develop a nonlinear theory for gel diffusiophoresis, followed by the presentation of our results in Section~\ref{sec:results}. We discuss our findings and conclude the paper in Section~\ref{sec:conclusion}.

\section{\label{sec:diffusiophoresis} Hydrogel diffusiophoresis}

Hydrogel diffusiophoresis is the deformation of a hydrogel to minimize the interaction energy between a solute gradient and the gel polymer network. Since it requires a solute gradient that leads to a gradient of pairwise interactions, its mechanism is different from the solubility of a polymer suspension or hydrogel at equilibrium. We proposed it to explain the transient osmotic swelling of PAA hydrogels upon acid-triggered copper or calcium release from the gel backbone where the hydrogel interface is fully permeable to the copper, calcium, and acid, in contrast with typical osmosis associated with interface selectivity to a solute~\cite{Katke_2024}. The equations for poroelastic gel deformations in the presence of gel diffusiophoresis suggest that tuning the interaction strength can dramatically speed up gel deformations, giving rise to superdiffusive dynamics that is otherwise diffusive.  

\begin{figure}[h]
            \centering
            \includegraphics[width=1\columnwidth]{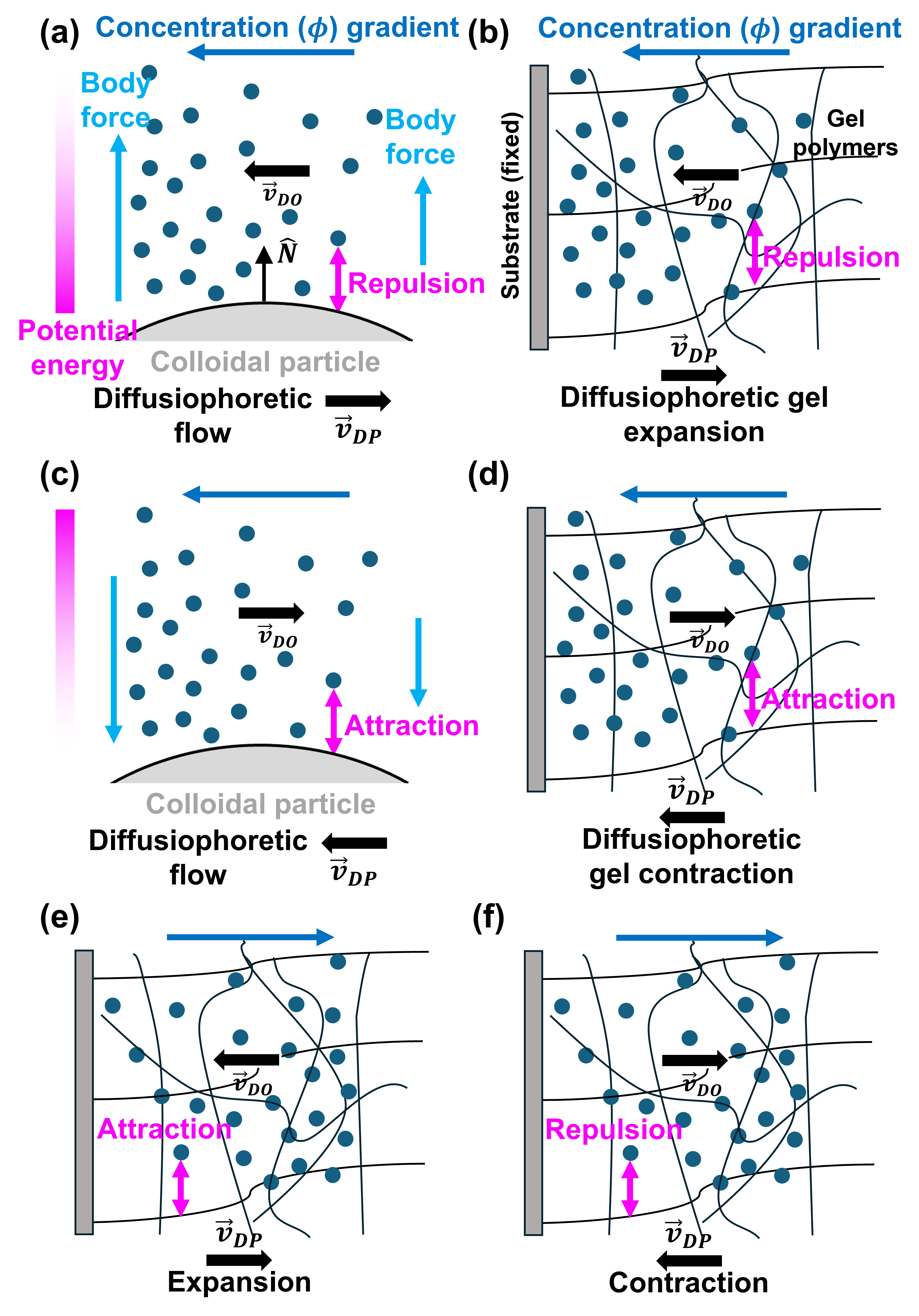}
            \caption{\textbf{Colloidal versus gel diffusiophoresis.} \textbf{(a), (c)} A colloidal particle (gray) with radius $R_c$ surrounded by a smaller particle species (dark green disks) with radius $r_s$ and volume fraction $\phi\,.$ A gradient $\nabla\phi$ is externally imposed. Blue arrows: body forces. Magenta color scale: potential energy variation for (a) pairwise repulsion, (c) pairwise attraction between the colloidal particle and small particles. Black arrows: Diffusio-osmotic flow of the solution (velocity: $\mathbf{v}_{DO}$) and counteracting diffusiophoretic flow of the colloidal particle (velocity: $\mathbf{v}_{DP}$). The normal vector to the colloidal particle surface $\mathbf{\hat{N}}$ points upward in the vertical direction when $R_c\gg r_s\,.$ \textbf{(b), (d), (e), (f)} Black lines: polymer network of a hydrogel. The hydrogel is adhered to a substrate (gray slab). Black arrows: Diffusio-osmotic flow of the solution (velocity: $\mathbf{v}_{DO}$) and counteracting diffusiophoretic flow of the gel (velocity: $\mathbf{v}_{DP}$). In (d)--(f), depending on the combination of the pairwise interactions (repulsion or attraction) and the gradient direction, diffusiophoretic gel expansion or contraction is generated.}
            \label{fig:diffusiophoresis}
        \end{figure}

Fig.~\ref{fig:diffusiophoresis} demonstrates gel diffusiophoresis in analogy with the well-known colloidal diffusiophoresis~\cite{DERJAGUIN1993138, Marbach2019}: In Fig.~\ref{fig:diffusiophoresis}a, a colloidal particle is surrounded by an externally maintained horizontal gradient of a smaller particle species (the solute) with a volume fraction $\phi$. For repulsive interactions between the colloidal particle surface and  the solute, the gradient of the interaction potential $\nabla U$ is antiparallel to the surface normal. Therefore, the local body force $-\phi\nabla U/v_c$ is parallel to the surface normal ($v_c:$ molecular volume of a solute particle). Per the Stokes equations, the variation of the upward body force yields a pressure drop along the surface towards higher force strength, driving a ``diffusio-osmotic flow" of the solution represented by the velocity $\mathbf{v}_{DO}$ in Fig.~\ref{fig:diffusiophoresis}a~\cite{DERJAGUIN1993138, Marbach2019}. Momentum conservation demands that the diffusio-osmotic flow be counteracted by the motion of the colloidal particle in the opposite direction with a velocity $\mathbf{v}_{DP}=-\mathbf{v}_{DO}\,,$ known as diffusiophoresis. Instead of a colloidal particle, when there is a hydrogel (a crosslinked polymer network) that interacts with the solute, henceforth called the "diffusio-osmotic agent," both the diffusio-osmotic and diffusiophoretic flows must be preserved. Importantly, the diffusiophoretic flow may lead to gel deformations because of the inherent elasticity of the gel. These deformations must especially be pronounced if the hydrogel is adhered to a substrate at its one end that imposes a no-displacement boundary condition (Fig.~\ref{fig:diffusiophoresis}b). If repulsive interactions are replaced by attractive interactions while the concentration gradient is left unchanged, all other arrows (body forces, diffusio-osmotic and diffusiophoretic flow velocities) change direction (Fig.~\ref{fig:diffusiophoresis}c). Then, the gel should not expand, but rather contract when $\nabla\phi$ is still to the left (Fig.~\ref{fig:diffusiophoresis}d). It suffices to switch the direction of the solute gradient again to induce expansion by attraction due to the symmetry broken by the no-displacement boundary condition at the lower gel boundary (Fig.~\ref{fig:diffusiophoresis}e). And while the solute gradient is to the right, the gel must now contract under repulsive interactions (Fig.~\ref{fig:diffusiophoresis}f). Figs.~\ref{fig:diffusiophoresis}b,~d--f demonstrate the difference of gel diffusiophoresis from regular enthalpic effects that determine solubility: The gradient direction can potentially generate swelling or contractile stress, irrespective of the sign of the polymer-solute interactions that lead to the equilibrium classification of ``good solvents," which would expand the gel due to pairwise attraction, versus ``poor solvents," which would contract the gel due to pairwise repulsion. For example, acetone as a poor solvent can collapse polyacrylamide hydrogels through a first-order volume phase transition~\cite{Tanaka1978, Tanaka1980}. In principle, when a gradient is applied across a gel, both the regular enthalpic effects as well as gel diffusiophoresis must be at play; they can compete as in Fig.~\ref{fig:diffusiophoresis}b,~d or reinforce each other as in Fig.~\ref{fig:diffusiophoresis}e,~f.  Note, however, that although typical diffusio-osmosis near an interface is a reversible process under $\nabla\phi\rightarrow-\nabla\phi$ due to the kinematic reversibility of the Stokes flow, hydrogel diffusiophoresis is highly nonlinear and not invariant under time reversal: Nonlinearity emerges from hyperelastic network deformations that change the pore size (coupling between two ``conjugate forces", i.e., pore pressure and elastic stress), whereas time reversal symmetry is broken by the hysteresis that dissipates heat during loading and unloading of a hydrogel. To this end, we previously computed hysteresis loops for hydrogel diffusiophoresis to measure the gel power output (i.e., dissipated heat)~\cite{Katke_2024}. For these reasons, Onsager reciprocal relations do not hold for hydrogel diffusiophoresis unlike typical diffusio-osmosis~\cite{PhysRev.37.405, PhysRev.38.2265, Ajdari2006, Marbach2019, Hinestrosa2020}.  Altogether, conversion of energy input (to maintain the solute gradient) to mechanical work can be harnessed to drive desired gel deformations by imposing appropriate external fields and boundary conditions, attributing an innate versatility to the gel diffusiophoresis.

\section{\label{sec:methods} Hyperelastic theory for gel diffusiophoresis}

Although we developed a linear poroelastic theory for small diffusiophoretic gel deformations in Ref.~\cite{Katke_2024}, speeding them up involves not only untapping superdiffusive scaling for shorter deformation times, but also achieving higher strains, which together yield enhanced strain rates without any structural modifications of the gel. To address the high strain limit, here we extend the theory of gel diffusiophoresis to large deformations by using the framework of nonlinear poroelasticity~\cite{biot1972theory, coussy2004poromechanics}. We apply this hyperelastic theory to two models: 
\begin{itemize}[leftmargin=*, itemsep=1pt, topsep=2pt, partopsep=1pt]
    \item In \textit{Model I}, an external gradient of the diffusio-osmotic agent is applied across the gel, and the gel deforms in response to it. This is the simplest hypothetical case, yet it is realizable in a microfluidic system that accommodates hydrogels to apply chemical gradients~\cite{Goy2019, Sleeboom2017, Beebe2000}. 
    \item In \textit{Model II}, a stored agent is released from within the gel upon a chemical trigger and diffuses outwards with a transient concentration gradient into a supernatant domain. The gel temporarily exhibits large deformations while a sufficiently steep interaction gradient persists. We envisage this model as a large deformation scenario for our chemically responsive divalent-ion-laden PAA gel system that reacts with acid~\cite{Korevaar2020, Katke_2024}. 
\end{itemize}

We take the swollen hydrogel (i.e., its ``wet" state) in equilibrium with a solvent as the reference configuration (or reference state) and adopt a Lagrangian (material) description of gel deformations. We employ a notation for tensor operations such that when $\mathbf{A}$ and $\mathbf{B}$ are rank-two tensors bundled from vectors in $\mathbb{R}^3$, $\mathbf{C}=\mathbf{A}\mathbf{B}$ is a rank-two tensor resulting from the contraction of $\mathbf{A}$ and $\mathbf{B}\,.$ We use the dot product ``$\cdot$" to denote contraction only when a vector is involved: Defining $\mathbf{d}\,, \mathbf{e}\in \mathbb{R}^3$ as vectors, $\mathbf{d}\cdot\mathbf{e}$ returns a scalar (the usual scalar product), $\mathbf{d}\cdot\mathbf{A}$ or $\mathbf{B}\cdot\mathbf{d}$ return a vector, whereas $\mathbf{d}\mathbf{e}$ returns a rank-two tensor. With this notation used henceforth, let $\mathbf{X}$ be position vector of the material point in the three-dimensional (3D) reference configuration and $\mathbf{x}(\mathbf{X},t)$ be the position vector of the material point in the current configuration. Then, the matrix displacement vector $\mathbf{u}$ and the deformation tensor $\mathbf{F}$ are defined by ($\nabla:$ gradient operator in the reference configuration):  
    \begin{equation}
        \mathbf{x} \equiv \mathbf{X} + \mathbf{u}(\mathbf{X},t)\,, \quad  \mathbf{F}\equiv \frac{d\mathbf{x}}{d\mathbf{X}}= \mathbf{I} + \nabla\mathbf{u}\,. 
        \label{eq:ch2_def_tensor}
    \end{equation}
    The Jacobian determinant $J$ measures the local volume change relative to the reference configuration,
    \begin{equation}
        J \equiv \det\mathbf{F}\,.
        \label{eq:ch2_detF}
    \end{equation}

The hydrogel composition requires us to develop a two-phase description to separately track the nominal volume fraction of the fluid phase $\Phi_f$ and that of the solid phase $\Phi_s$ in the material frame. In the lab frame, we denote the true volume fractions by the lower case $\phi$ with appropriate subscripts and superscripts; $\phi_f$ and $\phi_s$ are thus the fluid phase and solid phase true volume fractions. The volume fractions in the two frames are related by a factor of the Jacobian determinant $J$ (Eq.~\ref{eq:ch2_detF}) for all the phases and solutes as 
    \begin{equation}
        \phi = \frac{\Phi}{J}\,.
        \label{eq:ch2_relating_vol_frac_lab_mat}
    \end{equation} 
    
    To devise a nonlinear constitutive law for hydrogel mechanics under diffusiophoretic stress, we consider the poroelastic, enthalpic, and entropic contributions to the hydrogel stress tensor $\boldsymbol{\sigma}$ in the reference state, known as the nominal or First Piola-Kirchhoff stress. When $\boldsymbol{\sigma}$ is known, the true stress $\boldsymbol{\Sigma}$ (the stress tensor in the lab frame, also known as the Cauchy stress) can be calculated from the identity~\cite{hong2008, chaves2013notes}
    \begin{equation}
     \boldsymbol{\Sigma}=\boldsymbol{\sigma}\mathbf{F}/\det\mathbf{F}\,.   
     \label{eq:stress_conversion}
    \end{equation}
    
    Using the Flory free energy density for polymer network deformations ($\mu:$ shear modulus)~\cite{Flory1953}
    \begin{equation}
    W_f=\frac{\mu}{2}\left[ \Tr\left( \mathbf{F}^\top\cdot \mathbf{F}\right)-3-2\ln\det\left(\mathbf{F}\right)\right]\,,
        \label{eq:Flory}
    \end{equation}
    the poroelastic nominal stress tensor is given as ($\mathbf{I}:$ identity tensor, $p:$ interstitial solution or pore pressure)
    \begin{equation}
        \boldsymbol{\sigma}_{PE}=\frac{\partial W_f}{\partial \mathbf{F}}-p\mathbf{I}=\mu \left(\mathbf{F}- \mathbf{F}^{-\top}\right)-p\mathbf{I}\,.
        \label{eq:poroelastic_stress}
    \end{equation}
    
The entropic part $\boldsymbol{\sigma}_S$ comprises the entropy of mixing of the solid and fluid phases. The fluid phase of the hydrogel contains the liquid solvent with a volume fraction $\Phi_l$ and $n$ model-specific solute species with volume fractions $\Phi^{(0)}_\alpha$ such that $\Phi_f=\Phi_l+\sum_{\alpha=1}^n \Phi^{(0)}_\alpha\,.$ Similarly, the solid phase of the hydrogel is composed of the polymers with a volume fraction $\Phi_p$ and $m$ species with volume fractions $\Phi^{(b)}_\beta$ that can bind with the polymers or be released into the fluid phase, such that $\Phi_s=\Phi_p+\sum_{\beta=1}^m \Phi^{(b)}_\beta$ ($m\leq n$). The volume fractions $\Phi^{(0)}_\alpha$ and $\Phi^{(b)}_\beta$ are all zero in the reference state of the gel and will be explicitly defined for Models I and II in Sec.~\ref{subsec:Model1} and~\ref{subsec:Model2}. As we follow the deformation of the gel in the reference state, the polymer volume fraction remains constant at a value $\Phi_{p}=\Phi_{p,0}\,.$ Furthermore, volume conservation satisfies
    \begin{equation}
        J = \Phi_f + \Phi_s \,.
          \label{eq:ch2_volume_constraint}
    \end{equation} 
    In our formulation, we employ Eq.~\ref{eq:ch2_volume_constraint} separately rather than adding it to Eq.~\ref{eq:Flory} with a Lagrange multiplier, which would lead to an extra term in Eq.~\ref{eq:poroelastic_stress} (cf. Ref.~\cite{hong2008}). The two formulations are equivalent since the volume change upon changing the number of solvent or solute particles in the hydrogel is correctly accounted for either way.
    Using Eqs.~\ref{eq:stress_conversion} and~\ref{eq:ch2_volume_constraint}, $\boldsymbol{\sigma}_S$ is found from the mixing entropy of a dilute multi-component gel system when the diffusio-osmotic agent volume fraction is small and the agent is smaller than the pore size ($v_c\ll k_{f}^{3/2}$) such that it diffuses at the same rate as the solvent (thus not generating any osmotic stress in the gel to a leading order)~\cite{DoiSoftMatter}:
    \begin{equation}
    \boldsymbol{\sigma}_S=\frac{k_BT}{v_c}J\mathbf{F}^{-1}\left[\text{log}\left(1-\frac{\Phi_s}{J}\right)+\frac{\Phi_s}{J}\right]\,.
        \label{eq:entropic_stress}
    \end{equation}
    
    The enthalpic part of the stress tensor $\boldsymbol{\sigma}_H$ originates either from the conversion of a free solute particle into the bound state with finite $\Phi^{(b)}_\beta$ and decomplexation thereof into the fluid phase with finite $\Phi^{(0)}_\alpha$ or from polymer-solute interactions whose gradient induces diffusiophoresis when a solute gradient exists across the gel (Sec.~\ref{sec:diffusiophoresis}). We designate a tensor function $\boldsymbol{\mathcal{G}}(\mathbf{F}, \Phi^{(b)}_\beta)$ to the complexation induced stress contributions, to be defined for Models I and II later in Sec.~\ref{subsec:Model1} and~\ref{subsec:Model2}. For the stress induced by the polymer-solute interactions, which we will call ``diffusiophoretic stress," we assume that only one diffusio-osmotic agent is present in the system with a volume fraction $\Phi^{(0)}\equiv\Phi^{(0)}_{\alpha=1}.$ This assumption will be detailed for Models I and II in Sec.~\ref{subsec:Model1} and~\ref{subsec:Model2}. Unless the polymer-agent interactions are electrostatic, we will show that the scalar counterpart of a diffusiophoretic stress tensor $\boldsymbol{\sigma}_{DP}\equiv -(k_BT/v_c)\boldsymbol{\eta}_{DP}\Phi^{(0)}$ ($\boldsymbol{\eta_{DP}}:$ unitless diffusiophoretic coefficient tensor, $v_c:$ molecular volume) produces the correct diffusiophoretic velocity for uniaxial deformations when the momentum conservation of the fluid is combined with Darcy's law for porous flow in the gel. We take the origin of the diffusiophoretic stress to be steric repulsions between neutral (uncharged) polymers and the free diffusio-osmotic agent in the fluid phase of the gel. The reason that we introduce a tensor variable for diffusiophoresis is that the diffusiophoretic velocity components must depend on the corresponding pore dimension (average distance between polymers in that direction) under general deformations. With this stress term, the linear elastic theory showed excellent agreement with the experimental uniaxial swelling profiles of the copper-laden PAA gel upon acid delivery~\cite{Katke_2024}. Combining the complexation-induced stress contributions and the diffusiophoretic stress, $\boldsymbol{\sigma}_H$ is given as
    \begin{equation}
    \boldsymbol{\sigma}_H=\boldsymbol{\mathcal{G}}(\mathbf{F}, \Phi^{(b)}_\beta)+\boldsymbol{\sigma_{DP}}=\boldsymbol{\mathcal{G}}(\mathbf{F}, \Phi^{(b)}_\beta)-\frac{k_BT}{v_c}\boldsymbol{\eta}_{DP}\Phi^{(0)}\,,
        \label{eq:enthalpic_stress}
    \end{equation}
    The sum of Eqs.~\ref{eq:poroelastic_stress},~\ref{eq:entropic_stress},~\ref{eq:enthalpic_stress} gives a constitutive relation for the nominal stress tensor of the hydrogel system
    \begin{eqnarray}
        \boldsymbol{\sigma} & = &\boldsymbol{\sigma}_{PE}+\boldsymbol{\sigma}_S+\boldsymbol{\sigma}_H\nonumber
        \\&=&\mu(\mathbf{F}- \mathbf{F}^{-\top})-p\mathbf{I} +\boldsymbol{\mathcal{G}}(\mathbf{F}, \phi^{(b)}_\beta)\nonumber-\frac{k_BT}{v_c}\boldsymbol{\eta}_{DP}\Phi^{(0)}\\&&+\frac{k_BT}{v_c}J\mathbf{F}^{-1}\left[\text{log}\left(1-\frac{\Phi_s}{J}\right)+\frac{\Phi_s}{J}\right]\,,
        \label{eq:ch2_stress_tensor}
    \end{eqnarray}  
    which satisfies force balance in the reference configuration of the gel (Sec.~\ref{sec:ap_stress_mat_frame}):
    \begin{equation}
        \nabla\cdot\boldsymbol{\sigma} = 0\,.
        \label{eq:ch2_stress_balance}
    \end{equation}
    Using Eq.~\ref{eq:ch2_stress_tensor}, the true stress $\boldsymbol{\Sigma}$ (the stress tensor in the lab frame, also known as Cauchy stress) can be calculated from Eq.~\ref{eq:stress_conversion}. In the lab frame, Eq.~\ref{eq:ch2_stress_balance} is then rewritten as $\nabla_x\cdot\boldsymbol{\Sigma}=0\,,$ where  $\nabla_x$ is the gradient operator in that frame (Sec.~\ref{sec:ap_stress_mat_frame}).

    To couple the nonlinear hydrogel poroelasticity with solvent flow, solute advection and diffusion, we formulate the conservation laws for the fluid phase in the material frame, as opposed to their traditional treatment in the lab (Eulerian) frame. For momentum conservation, we employ Darcy's law in the material frame to relate the pore pressure $p$ to the relative flow velocity $\mathbf{V_f}$ with respect to the matrix displacement velocity as ($\mu_f:$ kinematic viscosity; Sec.~\ref{sec:ap_transport_mat_frame}) \cite{coussy_ch3, MacMinn2016} 
    \begin{equation}
        \Phi_f\mathbf{V_f}=-\frac{\mathbf{k}_f}{\mu_f}J \mathbf{F}^{-1}\mathbf{\mathbf{F}^{-\top}}\cdot\nabla p\,,
        \label{eq:ch2_darcy_flow}
    \end{equation}
    where $\mathbf{k}_f$ is the symmetric, positive-definite hydraulic permeability tensor. To derive the mass conservation law in the material frame, we start from the incompressibility condition in the lab frame, given by
    \begin{equation}
        \nabla\cdot(\phi_f \mathbf{v}_f+\phi_s\mathbf{v}_s)=0\,,
        \label{eq:incompressibility_lab}
    \end{equation}
    where $\mathbf{v}_f\,, \mathbf{v_s}\equiv \partial \mathbf{u}/dt$ are the velocities of the fluid phase and the polymer network (solid phase). Rewriting the incompressibility condition in the material frame gives (Sec.~\ref{sec:ap_transport_mat_frame})
    \begin{equation}
    \frac{d J}{dt}+\nabla\cdot\left(\Phi_f \mathbf{V}_f\right)=0\,,
        \label{eq:incompressibility}
    \end{equation}
    where $d/dt$ is the total time derivative operator in the material frame (cf. Eq.~\ref{eq:app_extra1}).

Eqs.~\ref{eq:ch2_def_tensor},~\ref{eq:ch2_detF},~\ref{eq:ch2_volume_constraint},~\ref{eq:ch2_stress_tensor},~\ref{eq:ch2_stress_balance},~\ref{eq:ch2_darcy_flow},~\ref{eq:incompressibility} constitute our nonlinear poroelastic theory for gel diffusiophoresis. These equations govern the time-dependent hydrogel deformations when the polymer-agent interactions (given, e.g., by Eq.~\ref{eq:ch_2_eta_DP}), the gel hydraulic permeability (given, e.g., by Eq.~\ref{eq:ch2_permeability}), and an equation governing the solid volume fraction $\Phi_s$ are determined. For a hydrogel whose polymer composition is intact, the latter equation is $\Phi_s=\Phi_{p, 0}\,.$ When solute species that interact with the polymers (either through diffusiophoresis or chemical reactions) are present, a kinetic equation for $\Phi_s$ needs to be introduced to consider the chemical conversion between $\Phi^{(0)}_\alpha$ and $\Phi^{(b)}_\beta$, as well as $m$ kinetic equations for $\Phi^{(b)}_\beta$ and $n$ advection-reaction-diffusion equations for $\Phi^{(0)}_\alpha\,.$ Then, defining appropriate initial and boundary conditions will fully determine the system of equations as detailed below for Models I and II. 

         \begin{figure*}
        \centering
        \includegraphics[width=1.0\textwidth]{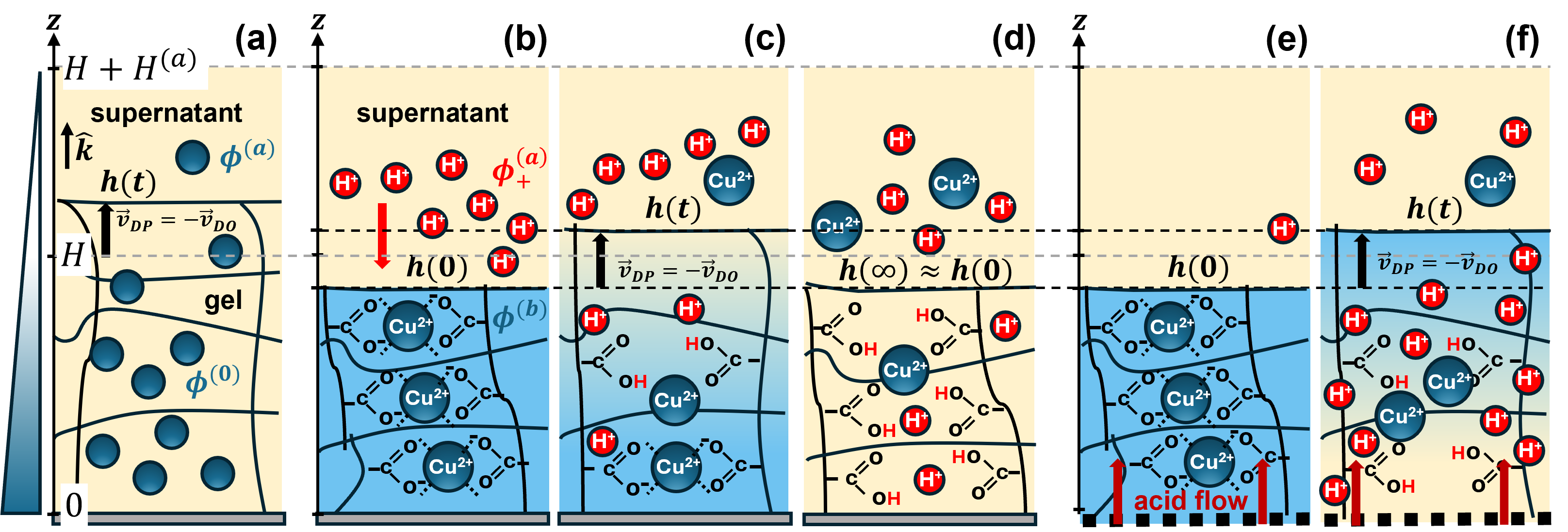}
        \caption{\textbf{Models I and II for gel diffusiophoresis.} \textbf{(a)} Model I: Gel response to an external solute gradient. A solute gradient (shown by the triangle with a dark blue color gradient) is maintained across a hydrogel, where the solute particles (dark blue) and the gel polymers interact through steric repulsion. In the frame of the gel, the interaction gradient generates diffusio-osmotic solvent flow from the supernatant domain with a velocity $\mathbf{v}_{DO}$ that is counteracted by the diffusiophoretic swelling of the gel with a velocity $\mathbf{v}_{DP}$ (Fig.~\ref{fig:diffusiophoresis}). \textbf{(b)}--\textbf{(d)} Model II, Scenarios 1 and 2: PAA gel response to competing stimuli. \textbf{(b)} Acid (red, volume fraction $\phi^{(a)}_+$) is delivered from the supernatant solution into a copper-laden PAA hydrogel attached to a substrate with a contracted initial height $h(0)<H$ due to the chelation between COO$^-$ and Cu$^{2+}$ (blue, volume fraction $\phi^{(b)}$), which turns the gel blue. \textbf{(c)} The formation of COOH groups (volume fraction $\phi^{(b)}_+$) releases Cu$^{2+}$ with a volume fraction $\phi^{(0)}$ into the gel solution. The gel swells with a time-dependent height $h(t)>h(0)$ and loses blue color while a gradient $\nabla\phi^{(0)}$ along the $-z$ axis emerges~\cite{Korevaar2020}. The diffusiophoretic swelling velocity $\mathbf{v}_{DP}$ negates the diffusio-osmotic solvent velocity $\mathbf{v}_{DO}$ (Fig.~\ref{fig:diffusiophoresis}). \textbf{(d)} The copper gradient, $\mathbf{v}_{DO}\,, \mathbf{v}_{DP}$ eventually vanish due to Cu$^{2+}$ diffusion, and the gel relaxes to the COOH-induced final height $h(\infty)\approx h(0)\,.$ The same dynamics is observed when Cu$^{2+}$ is replaced by the calcium ion Ca$^{2+}$~\cite{Korevaar2020}. \textbf{(e)},~\textbf{(f)} Model II, Scenario 3: PAA gel response to acid flow. \textbf{(e)} Acid flows into the hydrogel through the semipermeable substrate at $z=0\,.$ \textbf{(f)} Acid decomplexes the bound copper and releases it into the fluid phase of the gel. Due to the diffusiophoretic interactions between the gradient of free Cu$^{2+}$ and the polymer network, the hydrogel transiently swells and returns to the height imposed by the flow when the Cu$^{2+}$ gradient vanishes. Figure panels (b)--(d) are reproduced from Ref.~\cite{Katke_2024}.}
        \label{fig:ch2_acid_throughflow_schematic}
    \end{figure*}

\subsection{\label{subsec:Model1} Model I: Gel diffusiophoresis in an external solute gradient}

To construct a minimal model for gel diffusiophoresis, we assume that there exists only one solute species, which acts as a diffusio-osmotic agent. The polymer network and the agent solely interact through steric repulsion with an exclusion radius $R_e\,.$ The model involves no chemical reactions, leading to $\boldsymbol{\mathcal{G}}=0$ in Eq.~\ref{eq:ch2_stress_tensor}. Furthermore, the true agent volume fraction $\phi^{(0)}$ is held constant at both ends of the hydrogel-supernatant domain, such that it reaches a steady linear profile under diffusion along the domain when the gel stops deforming. The fixed boundary conditions for $\phi^{(0)}$ reduce the problem to 1D along the long axis of the domain in the absence of an external flow through the hydrogel (Fig.~\ref{fig:ch2_acid_throughflow_schematic}a). In this limit, the gel can only deform uniaxially when one of its ends is fixed. Therefore, the independent spatial variables herein are $Z\equiv|\mathbf{X}|$ in the material frame of the gel and $z\equiv |\mathbf{x}|$ in the supernatant domain in the lab frame. Furthermore, all vector and tensor variables reduce to scalars for uniaxial deformations: We denote the relative flow speed in the gel by $V_f(Z,t)\equiv|\mathbf{V}_f|$ (Eq.~\ref{eq:ch2_darcy_flow}), the nominal stress $\sigma_{ZZ}\equiv\mathbf{\hat{k}}\cdot\boldsymbol{\sigma}\cdot\mathbf{\hat{k}}\,,$ the scalar deformation $F_{ZZ}\equiv\mathbf{\hat{k}}\cdot\mathbf{F}\cdot\mathbf{\hat{k}}=1+\partial u_z/\partial Z=J$ equal to the local ``1D volume change," the scalar hydraulic permeability $k_f\equiv \mathbf{\hat{k}}\cdot\mathbf{k}_f\cdot\mathbf{\hat{k}}\,,$ the scalar diffusiophoretic coefficient $\eta_{DP}\equiv \mathbf{\hat{k}}\cdot\boldsymbol{\eta}_{DP}\cdot\mathbf{\hat{k}}\,,$ and $\mathbf{\hat{k}}\cdot\mathbf{I}\cdot\mathbf{\hat{k}}=1\,.$ 

We consider steric interactions between the diffusio-osmotic agent and the polymers as the source of gel diffusiophoresis as with our previous work~\cite{Katke_2024}. In this case, the diffusiophoretic stress $\sigma_{DP}$ has linear dependence on $\Phi^{(0)}$ as in Eq.~\ref{eq:enthalpic_stress}, and $\eta_{DP}$ is given by ($R_e:$ exclusion radius between the diffusio-osmotic agent and polymers)~\cite{Katke_2024} 
    \begin{equation}
        \eta_{DP} = \frac{R_e^2}{2k_f(\phi_s)}\,,
        \label{eq:ch_2_eta_DP}
    \end{equation}
    where the dependence of the gel hydraulic permeability $k_f$ on the solid phase volume fraction $\phi_s=\Phi_s/J\,,$ i.e., change in the pore size, is taken into account for large deformations. This implies an enhancement (suppression) of the diffusiophoretic effect with decreasing (increasing) pore area due to the change of the effective polymer surface density. Additionally, the volumetric scaling of the nominal volume fraction $\Phi_s=J \phi_s$ in the diffusioporetic stress (Eq.~\ref{eq:enthalpic_stress}) will lead to an increase (decrease) in pairwise polymer-agent interactions with local gel expansion (contraction). That way, Eqs.~\ref{eq:ch2_stress_tensor},~\ref{eq:ch_2_eta_DP} incorporate the two competing surface and volume effects as higher order contributions to the gel diffusiophoresis of steric origin. 

This simple uniaxial model is based on three differential equations for the vertical displacement function $u_z (Z, t)\equiv |\mathbf{u}|\,,$ the nominal agent volume fraction $\Phi^{(0)}(Z, t)\,,$ and the true agent volume fraction in the supernatant domain $\phi^{(a)}(z, t)\,.$ Eqs.\ref{eq:ch2_stress_tensor}-\ref{eq:incompressibility_lab} determine the uniaxial deformations as follows: Eq.~\ref{eq:ch2_stress_tensor},~\ref{eq:ch2_stress_balance} reduce to 
\begin{eqnarray}
\sigma&=&\mu\left(J-J^{-1}\right)-p\nonumber\\&&-\frac{k_BT}{v_c}\left[\eta_{DP}\Phi^{(0)}-\text{log}\left(1-\frac{\Phi_s}{J}\right)-\frac{\Phi_s}{J}\right]\,,
\label{eq:stress_tensor_1D}
\end{eqnarray}
\begin{equation}
    \frac{\partial\sigma}{\partial Z}=0\,, 
    \label{eq:stress_balance_1D}
\end{equation}
which enforces constant stress across the gel. We thus set $\sigma=0$ that corresponds to the equilibrium wet state of the hydrogel with pure solvent in it. In Eq.~\ref{eq:stress_tensor_1D}, the solid phase is composed only of polymers, i.e., $\Phi_s=\Phi_{p,0}$ without any bound species. In terms of the material frame variables, the incompressibility condition in the lab frame (Eq.~\ref{eq:incompressibility_lab}) reduces to 
\begin{equation}
\Phi_f V_f+\frac{d u_z}{d t}=0    
\label{eq:incompressibility_1D}
\end{equation}
when there is no flow in the supernatant domain (cf. Eq.~\ref{eq:app_flux_continuity}). In 1D, Darcy's law in the material frame (Eq.~\ref{eq:ch2_darcy_flow}) becomes
\begin{equation}
\Phi_f V_f=-\frac{k_f}{\mu_f}\frac{1}{J}\frac{\partial p}{\partial Z}\,.
    \label{eq:Darcy_flow_1D}
\end{equation}
In accordance with the effect of a changing pore area on diffusiophoresis, we use a phenomenological description of $k_f$ by employing the functional form~\cite{Costa2006}
    \begin{equation}
        k_f = k_{f,0}\left[b\tanh{\left(a\frac{1-\Phi_s/J}{\Phi_s/J}\right)}\right]\,.
        \label{eq:ch2_permeability}
    \end{equation}
    The coefficients $a$, $b$ are chosen such that $\lim_{\mathbf{\Phi}_s\to 0}  k_f= H^2$ where $H$ is the shortest dimension of the hydrogel film (Stokes flow limit), and $k_f=k_{f,0}$ when $\Phi_s=\Phi_{p,0}\,,$ $k_{f,0}$ being the permeability in the reference state (Sec.~\ref{sec:ap_estimation_para_nl}). 
    
We define the equilibrium gel height in its wet state by $H$ and introduce the unitless variables $u_z'\equiv u_z/H\,, Z'\equiv Z/H$, $ t'\equiv t/\tau\,,$ $V'_f\equiv V_f\tau/H\,,$ where $\tau\equiv\mu_f H^2/k_{f, 0} \bar{p}$ is the poroelastic deformation timescale, $\bar{p}\equiv 2\mu$ is the pressure scale of the system, $k_f'=k_f/k_{f, 0}$ (Eq.~\ref{eq:ch2_permeability}). Dropping the primes, Eqs.~\ref{eq:stress_tensor_1D}--\ref{eq:Darcy_flow_1D} yield a dimensionless evolution equation for the gel displacement as ($\nu_{DP}\equiv k_B T \eta_{DP}/v_c\bar{p}$)~\cite{si_dimensionless_eq}
\begin{eqnarray}
\frac{\partial u_z}{\partial t}&=&\frac{k_f}{J}\frac{\partial}{\partial Z}\left\{\frac{1}{2}\left(J-J^{-1}\right)-\nu_{DP}\Phi^{(0)}\right.\nonumber\\&&\left.+\frac{k_B T}{v_c \bar{p}}\left[\text{log}\left(1-\frac{\Phi_{p, 0}}{J}\right)+\frac{\Phi_{p, 0}}{J}\right]\right\}\,,
    \label{eq:gel_displacement_model1}
\end{eqnarray}
The reason behind the definition $\bar{p}\equiv 2\mu$ is that when $\partial u_z/\partial Z\ll1\,,$ $(J-J^{-1})/2\approx \partial u_z/\partial Z$ in the linear elastic limit of Eq.~\ref{eq:gel_displacement_model1}, which then reduces to the unitless linear poroelastic evolution equation derived in Ref.~\cite{Katke_2024}. 

To solve for the displacement $u_z\,,$ Eq.~\ref{eq:gel_displacement_model1} must be complemented with a second equation for $\Phi^{(0)}(Z, t)\,.$ Noting that $d\Phi_{p}/dt=d\Phi_{p,0}/dt=0$ in the absence of bound species in the material frame, we modify Eq.~\ref{eq:incompressibility} to write a continuity equation for the agent as the subphase of the gel fluid. Since the agent with the volume fraction $\Phi^{(0)}(Z, t)$ is also subject to diffusion, the pertaining dimensionless continuity equation is given as ($\xi^{(0)}\equiv D^{(0)} \tau/H^2\,:$ unitless diffusivity, $D^{(0)}:$ diffusivity of the agent inside the gel)
\begin{equation}
    \frac{d \Phi^{(0)}}{dt}+\frac{\partial }{\partial Z}\overbrace{\left[\Phi^{(0)} V_f-\xi^{(0)}\frac{1}{J}\frac{\partial}{\partial Z}\left(\frac{\Phi^{(0)}}{J}\right)\right]}^{\equiv Q^{(0)}}=0\,.
        \label{eq:agent_continuity_model1}
    \end{equation}
    
The agent that overflows to the supernatant domain is only subject to diffusion. Defining the unitless lab frame coordinate as $z\equiv H^{(a)} z'\,,$ where $z=H^{(a)}$ is the position of the upper boundary of the supernatant domain, the unitless diffusivity as $\xi^{(a)}\equiv D^{(a)}\tau/\alpha^2 H^2\,,$ ($D^{(a)}:$ diffusivity of the agent in the supernatant domain, $\alpha\equiv H^{(a)}/H$), and dropping the primes, the unitless diffusion equation in the lab frame is given as

    \begin{equation}
    \frac{\partial \phi^{(a)}}{\partial t}=\frac{\partial}{\partial z}\overbrace{\left[\xi^{(a)}\frac{\partial \phi^{(a)}}{\partial z}\right]}^{\equiv Q^{(a)}}\,.
        \label{eq:agent_diffusion_model1}
    \end{equation}

Eqs.~\ref{eq:gel_displacement_model1}--\ref{eq:agent_diffusion_model1} are each second order in space and first order in time in $u_z\,, \Phi^{(0)},$ and $\phi^{(a)},$ respectively, thus requiring six boundary conditions and three initial conditions to fully determine Model I. Before we list them, we note that the stress balance at the gel-supernatant interface as a continuity condition necessitates a ``jump condition" between the hydrogel pore pressure $p$ and supernatant fluid pressure $P\,,$ which we specify as
\begin{equation}
    (P-p)\bigg|_{Z=H}=\frac{k_BT}{v_c}\left[\eta_{DP}\Phi^{(0)}-\text{log}\left(1-\frac{\Phi_s}{J}\right)-\frac{\Phi_s}{J}\right]\bigg|_{Z=H}\,,
    \label{eq:pressure_jump_model1}
\end{equation}
where $\Phi_s=\Phi_{p, 0}\,.$ We have taken Eq.~\ref{eq:pressure_jump_model1} from Ref~\cite{Katke_2024}: In the absence of any solute, $P-p$ is set by the osmotic pressure of the polymer network. The first and second terms in the osmotic pressure. i.e., the right-hand side of Eq.~\ref{eq:pressure_jump_model1}, are the enthalpic contribution from the agent and the contribution from the mixing entropy of the solid and fluid phases, respectively. The enhalpic contribution ensures that adding a diffusio-osmotic agent with a volume fraction $\Phi^{(0)}$ reduces the pore pressure $p$ with respect to the external pressure $P\,,$ i.e., decreases the local water content for repulsive interactions ($\eta_{DP}>0$)~\cite{hong2008, hong2009}. Thus, the pressure jump across the interface must depend both on the solid phase volume fraction $\Phi_s$ and the agent volume fraction $\Phi^{(0)}\,,$ leading to Eq.~\ref{eq:pressure_jump_model1}, which ensures chemical potential equilibration across the interface. Then, the stress balance condition at $Z=H$ based on Eqs.~\ref{eq:stress_tensor_1D}, ~\ref{eq:pressure_jump_model1}, and $P=0$ (since there is no net flow in the domain), the volume fraction and flux continuity conditions at $Z=H\,,$ the fixed boundary conditions for the gel displacement and inlet agent volume fraction at $Z=0\,,$ and the fixed boundary condition for outlet agent volume fraction at $Z=H+H^{(a)}$ are given in the unitless form as
    \begin{equation}
    \begin{split}
        & Z=0:\quad u_z=0\,,\quad \frac{\Phi}{J}=\phi_{in}\,,
        \\& Z=1:\quad \frac{1}{2}\left(J-J^{-1}\right)=0\,,\quad \frac{\Phi^{(0)}}{J}=\phi^{(a)}\,,
        \\&\hspace{4.5em} Q^{(0)}+\phi^{(0)}\frac{d u_z}{dt}=Q^{(a)}\,, 
        \\& z=1+\alpha:\quad \phi^{a}=\phi_{out}\,.
    \end{split}
    \label{eq:BC_model1}
    \end{equation}
At $Z=1$ in Eq.~\ref{eq:BC_model1}, stress balance yields the first condition, i.e., sudden elastic relaxation induced by the chemical potential equilibration (Eq.~\ref{eq:pressure_jump_model1}), the second condition is the volume fraction continuity in the lab frame when the gel interface is not selective to the solute, and the third condition is the flux continuity in the lab frame.

Taking the initial conditions as zero matrix displacement that corresponds to the equilibrium wet state of the gel and an externally imposed linear volume fraction profile of the agent between the inlet and outlet 
    \begin{eqnarray}
        && u_z(Z, t=0)=0\,,\quad \Phi^{(0)}(Z,0)=\frac{(\phi_{out}-\phi_{in})}{(1+\alpha)}Z+\phi_{in}\,,\nonumber\\&& \phi^{(a)}(z, 0)=\frac{(\phi_{out}-\phi_{in})}{ (1+\alpha)}\left[\alpha (z-1)+1\right]+\phi_{in}
        \label{eq:IC_model1}
    \end{eqnarray}
completes the formulation of Model I.

\subsection{\label{subsec:Model2} Model II: Gel diffusiophoresis through an internally released dynamic solute gradient} 

Here we apply our hyperelastic theory for gel diffusiophoresis to the swelling dynamics of copper-laden PAA hydrogel system. Our goal is to investigate how the strain rates can be amplified beyond the linear elastic regime: For reference, the measured swelling strain rate upon addition of 1 molar (1M) HCl (corresponding to 1M H$^+$) in Refs.~\cite{Korevaar2020, Katke_2024} corresponds to $\sim 0.4\tau^{-1}$ for the parameters used in this work (Table~\ref{table:simulation_parameters}). To induce bigger deformations, we consider three distinct scenarios:

\begin{enumerate}[topsep=2pt, partopsep=1pt, itemsep=1pt, leftmargin=*, itemindent=\leftmargin, label=\textit{Scenario \arabic*.}]
    \item Higher acid molarity (up to $\sim$5M acid): We will add higher volume fractions of acid on top of the hydrogel layer that is fixed on an impermeable rigid substrate (Fig.~\ref{fig:ch2_acid_throughflow_schematic}b--d).
     \item Changing the ionic radius of the agent: In the same simulation setup as in \textit{Scenario 1}, changing the agent size under constant acid molarity will tune the repulsive interactions and in turn the diffusiophoretic stress (Eqs.~\ref{eq:enthalpic_stress},~\ref{eq:ch_2_eta_DP}).
    \item Flow through the hydrogel in a straight channel to increase the rate of acid delivery: We consider that the gel is fixed on a rigid semipermeable membrane that admits acid flow but blocks agent outflux (Fig.~\ref{fig:ch2_acid_throughflow_schematic}e, f). The flow rate needs to be lower than the flow scale $\sim k_{f, 0}\mu /(\mu_f H)$ that the hydrogel admits ($H:$ hydrogel layer thickness). Beyond this threshold rate, the hydrogel will behave as a porous plug.
   
\end{enumerate}
Since an internal gradient can be generated in any hydrogel geometry for arbitrary deformations, we first present the general model in 3D. Then, we reduce the equations to 1D to simulate the uniaxial gel deformations for the three scenarios.

\subsubsection{\label{subsubsec:Model2_3D} Equations of motion in 3D}

The origin of the PAA hydrogel diffusiophoresis is the gradient of the divalent ions Cu$^{2+}$ or Ca$^{2+}$~\cite{Katke_2024}. The solid phase with a volume fraction $\Phi_p$ is contracted while storing the agent with a bound volume fraction $\Phi^{(b)}(\mathbf{X}, t)\,,$ which is displaced by acid present in the gel fluid phase with a volume fraction $\Phi^{(0)}_+(\mathbf{X}, t)$ (Fig.~\ref{fig:ch2_acid_throughflow_schematic}b, c). The acid instantaneously makes bonds with the carboxylate groups on the polymers turning the gel hydrophobic, i.e. COO$^-\rightarrow$~COOH, with a bound volume fraction $\Phi^{(b)}_+(\mathbf{X}, t).$ The released agent acquires a volume fraction $\Phi^{(0)}(\mathbf{X}, t)$ in the gel fluid phase. That is, the set of solute and bound species have the volume fractions $\Phi^{(0)}_\alpha\equiv\{\Phi^{0}, \Phi^{0}_+\}\,,$ and $\Phi^{(b)}_\beta\equiv\{\Phi^{b}, \Phi^{b}_+\}\,,$ respectively. The diffusiophoretic gel swelling between the initial and final contracted states is driven by the steric interactions between the copper ions neutralized by a hydration shell and the polymers with neutral COOH groups until the gradient $\nabla\Phi^{(0)}$ diminishes while the copper ions diffuse from the gel to the supernatant domain (Fig.~\ref{fig:ch2_acid_throughflow_schematic}c, d).


To apply the hyperelastic theory to this system, we first determine the enthalpic stress tensor function $\boldsymbol{\mathcal{G}}(\mathbf{F}, \Phi^{(b)}, \Phi^{(b)}_+)$ in Eq.~\ref{eq:enthalpic_stress} by considering the nonlinear contractile stresses arising from the complexation of the agent and protons (acid) to the gel polymer network. Defining $\tilde{\chi}\,,\tilde{\gamma}>0$ as the stress prefactors associated with the complexation of the acid and agent, respectively, we employ the phenomenological relation (cf. Ref.~\cite{Katke_2024})
\begin{equation}
\boldsymbol{\mathcal{G}}\equiv\tilde{\chi}\Phi^{(b)}_+\mathbf{F}+\tilde{\gamma}\Phi^{(b)}\mathbf{F}\,.
    \label{eq:model2_enthalpic_stress}
\end{equation}
We will use the experimental initial (fully copper-complexed) and final (fully acid-complexed) equilibrium height measurements of the PAA hydrogel to determine the numerical values of $\tilde{\chi}$ and $\tilde{\gamma}$ (Sec.~\ref{sec:ap_estimation_para_nl}). 

Next, we split the incompressibility condition given by Eq.~\ref{eq:incompressibility} in the material frame into two continuity equations: one for the solid phase with the volume fraction $\Phi_s=\Phi_p+\Phi^{(b)}+\Phi^{(b)}_+$ and another for the fluid phase with the volume fraction $\Phi_f=\Phi_l+\Phi^{(0)}+\Phi^{(0)}_+.$ This allows us to take into account the conversion rates $R_{Cu}$ and $R_+$ between the particles in the fluid phase ($\Phi^{(0)}$ and $\Phi^{(0)}_+$) and their bound-state counterparts in the solid phase ($\Phi^{(b)}$ and $\Phi^{(b)}_+$). The two continuity equations are given as (Sec.~\ref{sec:ap_transport_mat_frame})
    \begin{equation}
        \frac{d\Phi_s}{d t} =  -R_{Cu}+R_+\,,      
        \label{eq:ch2_solid_mass_cons}
    \end{equation}
    \begin{equation}
        \frac{d\Phi_f}{d t} + \nabla\cdot(\Phi_f\mathbf{V_f}) =  R_{Cu}-R_+\,.      
        \label{eq:ch2_fluid_mass_cons}
    \end{equation}
    
    Each of the free and bound volume fractions are governed by the following conservation laws in the material frame ($D_{Cu}:$ Cu$^{2+}$ diffusivity in the gel, $D_+:$ acid diffusivity in the gel, $\Phi^{\ast}:$ fixed volume fraction of the COO$^-$ groups on the PAA polymer network): 
    \begin{eqnarray} 
        \frac{d\Phi^{(0)}}{d t}&+& \nabla\cdot\overbrace{\left[\Phi^{(0)}\mathbf{V_f} - D_{Cu}J\mathbf{F}^{-1}\mathbf{F}^{-\top}\cdot\nabla\left(\frac{\Phi^{(0)}}{J}\right)\right]}^{\equiv \mathbf{Q}_{Cu}}\nonumber\\&=&\underbrace{\frac{1}{J}\left[\tilde{r}\Phi^{(0)}_+\Phi^{(b)}-\tilde{r}\Phi^{(0)}(\Phi^*-2\Phi^{(b)}-\Phi^{(b)}_+)\right]}_{\equiv R_{Cu}}\,,
        \label{eq:ch2_free_copper_gel}
    \end{eqnarray}
    \begin{eqnarray}
         \frac{d\Phi^{(0)}_+}{d t} &+& \nabla\cdot\overbrace{\left(\Phi^{(0)}_+\mathbf{V_f} - D_{+}J\mathbf{F}^{-1}\mathbf{F}^{-\top}\cdot\nabla\left(\frac{\Phi^{(0)}_+}{J}\right)\right)}^{\equiv \mathbf{Q}_{+}}\nonumber\\&=& -\underbrace{\frac{1}{J}\left[\tilde{r}\Phi^{(0)}_+(\Phi^*-\Phi^{(b)}_+)\right]}_{\equiv R_+}\,,
        \label{eq:ch2_free_acid_gel}
    \end{eqnarray}
    \begin{equation}
        \frac{d\Phi^{(b)}}{d t} = -R_{Cu}\,,
        \label{eq:ch2_bound_copper_gel}
    \end{equation}
    \begin{equation}
        \frac{d\Phi^{(b)}_+}{d t} = R_{+}\,.
        \label{eq:ch2_bound_acid_gel}
    \end{equation}
   The fluxes $\mathbf{Q}_{Cu}$ and $\mathbf{Q}_+$ in Eqs.~\ref{eq:ch2_free_copper_gel} and~\ref{eq:ch2_free_acid_gel} include advection and diffusion terms, while $R_{Cu}$ and $R_+$ are the source terms also appearing in Eqs.~\ref{eq:ch2_solid_mass_cons},~\ref{eq:ch2_fluid_mass_cons}. The first term of $R_{Cu}$ represents the acid-induced Cu$^{2+}$ release from the gel backbone, and the second term is the formation rate of a COO$^--$Cu$^{2+}-$COO$^-$ chelate. The source term $R_+$ is the COOH formation rate. The derivations of these equations are given in Sec.~\ref{sec:ap_transport_mat_frame}.

    To model the flow in the supernatant domain, we work in the lab frame. Let $\mathbf{V}(\mathbf{x})$ be the fluid velocity in the supernatant domain and $P$ be the fluid pressure. Then, the fluid stress tensor is given by
    \begin{equation}
        \quad \boldsymbol{\sigma}^{(a)} = \mu_f \nabla \mathbf{V}- \mathbf{I} P\,. 
        \label{eq:ch2_fluid_trans_supernatant}
    \end{equation}
    In the low Reynolds number regime, the flow is governed by the Stokes flow equation and the incompressibility condition:
    \begin{equation}
        \quad \nabla\cdot\boldsymbol{\sigma}^{(a)} = 0\,, \quad \nabla\cdot\mathbf{V}=0\,.
        \label{eq:ch2_stokes flow}
    \end{equation}
    
    The dynamics of the free copper and free acid in the supernatant domain are solely governed by advection and diffusion. Denoting the volume fractions of free copper and free acid in the supernatant in the lab frame by $\phi^{(a)}$ and $\phi^{(a)}_+\,,$ respectively, and their corresponding diffusivities by $D_{Cu}^{(a)}$ and $D_+^{(a)}\,,$ the pertinent continuity equations are given by
    
    \begin{equation}
        \frac{\partial \phi^{(a)}}{\partial t} + \nabla\cdot(\overbrace{\phi^{(a)}\mathbf{V} - D_{Cu}^{(a)}\nabla \phi^{(a)}}^{\mathbf{Q}_{Cu}^{(a)}}) = 0\,,
        \label{eq:ch2_free_copper_supernatant}
    \end{equation}

    \begin{equation}
        \frac{\partial \phi^{(a)}_+}{\partial t} + \nabla\cdot(\overbrace{\phi^{(a)}_+\mathbf{V} - D_{+}^{(a)}\nabla \phi^{(a)}_+}^{\mathbf{Q}_{+}^{(a)}}) = 0\,.
        \label{eq:ch2_free_acid_supernatant}
    \end{equation}

 Eqs.~\ref{eq:ch2_def_tensor},~\ref{eq:ch2_detF},~\ref{eq:ch2_volume_constraint},~\ref{eq:ch2_stress_tensor},~\ref{eq:ch2_stress_balance},~\ref{eq:ch2_darcy_flow},~\ref{eq:model2_enthalpic_stress}--\ref{eq:ch2_free_acid_supernatant} constitute a 3D general theory for the diffusiophoretic deformations of an ion-laden hydrogel in response to an acid trigger. As a special case, we consider a hydrogel film adhered to a substrate and exposed to acid vertically from top and specify the boundary conditions accordingly  (Fig.~\ref{fig:ch2_acid_throughflow_schematic}b--f). We denote the vertical material coordinate along the hydrogel-supernatant domain by $Z\,.$ At the gel-supernatant boundary  $(Z=H)$, we impose continuity of the total mass flux and stresses in the lab frame. Furthermore, in accordance with our experimental system (Fig.~\ref{fig:ch2_acid_throughflow_schematic}b--f)~\cite{Korevaar2020, Katke_2024}, we impose volume fraction continuity for the free copper and the free acid at $Z=H$ (since the gel interface is fully permeable to both) as well as continuity in their fluxes evaluated in the lab frame. Defining $\mathbf{\hat{n}}$ as the unit normal of any boundary in this system with its vertical component in the $+Z$ direction, the continuity conditions at the gel-supernatant interface at $Z=H$ are given by (Sec.~\ref{sec:ap_transport_mat_frame})    
    \begin{eqnarray}
        &&\frac{\mathbf{F}}{J}\cdot\left(\Phi_f\mathbf{V_f}\right) + \frac{\partial \mathbf{u}}{\partial t} = \mathbf{V}\,, \quad
        \left(\frac{1}{J}\boldsymbol{\sigma}  F^{\top}\right)\cdot\mathbf{\mathbf{\hat{n}}} = \boldsymbol{\sigma}^{(a)}\cdot\mathbf{\hat{n}}\,, \nonumber
        \\&&P-p=\frac{k_B T}{v_c} \left\{\left(\mathbf{\hat{n}}\cdot\boldsymbol{\eta}_{DP}\cdot\mathbf{\hat{n}}\right)\Phi^{(0)}\right.\nonumber\\&&\qquad\qquad \left.-J (\mathbf{\hat{n}}\cdot\mathbf{F}^{-1}\cdot\mathbf{\hat{n}})\left[\log\left(1-\frac{\Phi_s}{J}\right)+\frac{\Phi_s}{J}\right]\right\}\,,\nonumber
        \\&&\phi^{(a)} = \frac{\Phi^{(0)}}{J}\,,\quad 
        \phi^{(a)}_+ = \frac{\Phi^{(0)}_+}{J}\,,\nonumber
        \\&&\mathbf{Q}_{Cu}^{(a)}\cdot\mathbf{\hat{n}} = \left(\frac{\mathbf{F}}{J}\cdot \mathbf{Q}_{Cu}+\phi^{(0)}\frac{\partial \mathbf{u}}{\partial t}\right)\cdot\mathbf{\hat{n}}\,,\nonumber
        \\&&\mathbf{Q}_{+}^{(a)}\cdot\mathbf{\hat{n}} = \left(\frac{\mathbf{F}}{J}\cdot \mathbf{Q}_{+}+\phi^{(0)}_+\frac{\partial \mathbf{u}}{\partial t}\right)\cdot\mathbf{\hat{n}}\,.
        \label{eq:ch2_interface_bc} 
    \end{eqnarray}
    We specify the boundary conditions at $Z=0$ and $Z=H+H^{(a)}\,,$ where $H^{(a)}$ is the upper boundary of the supernatant domain, for each scenario as follows: Scenarios 1 and 2 occur between impermeable rigid boundaries that impose no-flux conditions, whereas Scenario 3 must admit the advection of acid through a steady inlet flow into the gel at $Z=0$ and outlet flow of the solvent at $Z=H+H^{(a)}$ because of incompressibility (Eq.~\ref{eq:ch2_stokes flow}). Since the hydrogel is adhered to the rigid substrate, the displacement vector $\mathbf{u}$ is zero at $Z=0\,.$ These extra boundary conditions complete the 3D poroelastic model for gel diffusiophoresis:
    \begin{equation}
    \begin{split}
        Z=0:&\quad \mathbf{u}=0\,,
        \quad \mathbf{\hat{n}}\cdot\mathbf{V} = 
        \begin{cases}
        0\,, & \textit{Scenarios 1, 2\,,}\\ 
        \mathbf{\hat{n}\cdot U} & \textit{Scenario 3\,,}
        \end{cases} 
        \\& \mathbf{\hat{n}}\cdot\left(\frac{\mathbf{F}}{J}\cdot \mathbf{Q}_{Cu}\right) =0\,,
        \\&\mathbf{\hat{n}}\cdot\left(\frac{\mathbf{F}}{J}\cdot \mathbf{Q}_{+}\right) = 
        \begin{cases}
        0\,, & \textit{Scenarios 1, 2\,,}\\ 
        \phi_+^{(f)}\mathbf{\hat{n}\cdot U} & \textit{Scenario 3\,,}
        \end{cases}
    \end{split}
    \label{eq:ch2_bottom_bc}
    \end{equation}
    and 
    \begin{equation}
    \begin{split}
        Z=H+H^{(a)}:&\; \mathbf{\hat{n}}\cdot\mathbf{V} = 
       \begin{cases}
        0\,, & \textit{Scenarios 1, 2\,,}\\ 
        \mathbf{\hat{n}\cdot U} & \textit{Scenario 3\,,}
        \end{cases}   
        \\& P =0\,, 
        \\& \mathbf{\hat{n}}\cdot\mathbf{Q}_{Cu}^{(a)} =0\,,
        \\& \mathbf{\hat{n}}\cdot\mathbf{Q}_{+}^{(a)} = 0\,.
    \end{split}
    \label{eq:ch2_top_bc}
    \end{equation}

    \subsubsection{\label{subsubsec:Model2_1D} Equations of motion in 1D}

For uniaxial deformations, we extend the formulation presented in Sec.~\ref{subsec:Model1} to the nonlinear swelling of copper-laden PAA hydrogel initiated by acid addition. Per Eqs.~\ref{eq:model2_enthalpic_stress} and~\ref{eq:ch2_interface_bc}-\ref{eq:ch2_top_bc}, Eq.~\ref{eq:stress_tensor_1D} and~\ref{eq:incompressibility_1D} are respectively modified to ($\tilde{U}\equiv |\mathbf{U}|$)
    \begin{eqnarray}
    \sigma&=&\mu\left(J-J^{-1}\right)-p+J\left(\tilde{\chi}\Phi^{(b)}_++\tilde{\gamma}\Phi^{(b)}\right)\nonumber\\&&-\frac{k_BT}{v_c}\left[\eta_{DP}\Phi^{(0)}-\text{log}\left(1-\frac{\Phi_s}{J}\right)-\frac{\Phi_s}{J}\right]\,,
    \label{eq:stress_tensor_model2}
    \end{eqnarray}
    \begin{equation}
    \Phi_f V_f+\frac{d u_z}{d t}=
    \begin{cases}
        0 & \textit{Scenario 1, 2\,,}\\
        \tilde{U} & \textit{Scenario 3\,.}
    \end{cases}    
    \label{eq:incompressibility_model2}
    \end{equation}
For \textit{Scenario 3}, we assume that the gel is subject to a constant pressure drop $\Delta \tilde{p}\,,$ and the flow speed is set by the minimum hydraulic permeability of the gel $\tilde{k}_{f, min}\,.$ We implement this in our simulations by using a minimum operator that finds the minimum permeability across the gel as a function of time.  This allows us to model the gel as a porous plug that regulates the flow as a function of its minimum permeability across its 1D structure. The external speed $\tilde{U}$ is then given by
\begin{equation}
  \tilde{U}=-\frac{\tilde{k}_{f, min}}{\mu_f}\frac{\Delta \tilde{p}}{H}\,.
  \label{eq:external_speed}
\end{equation}

The combination of Eq.~\ref{eq:stress_balance_1D},~\ref{eq:Darcy_flow_1D},~\ref{eq:stress_tensor_model2}, and~\ref{eq:incompressibility_model2} leads to the dimensionless evolution equation for the gel displacement in Model II ($\chi\equiv\tilde{\chi}/\bar{p}\,, \gamma\equiv\tilde{\gamma}/\bar{p}\,, k_{f, min}\equiv\tilde{k}_{f, min}/k_{f, 0}\,, \Delta p\equiv\Delta\tilde{p}/\bar{p}\,,$)
\begin{eqnarray}
\frac{\partial u_z}{\partial t}&=&U+\frac{k_f}{J}\frac{\partial}{\partial Z}\biggl\{\frac{1}{2}\left(J-J^{-1}\right)+J\left(\chi\Phi^{(b)}_++\gamma\Phi^{(b)}\right)\nonumber\\&&-\nu_{DP}\Phi^{(0)}+\frac{k_B T}{v_c \bar{p}}\left[\text{log}\left(1-\frac{\Phi_s}{J}\right)+\frac{\Phi_s}{J}\right]\biggr\}\,,
\label{eq:gel_displacement_model2}
\end{eqnarray}
\begin{eqnarray}
U&\equiv&-k_{f, min}\Delta p\,\begin{cases}
    =0 & \textit{Scenario 1, 2\,,}\\
    >0 & \textit{Scenario 3\,.}
\end{cases}
\label{eq:flow}
\end{eqnarray}

To complement Eqs.~\ref{eq:incompressibility_model2} and~\ref{eq:gel_displacement_model2}, the reduction of Eq.~\ref{eq:ch2_free_copper_gel}-~\ref{eq:ch2_bound_acid_gel},~\ref{eq:ch2_free_copper_supernatant},~\ref{eq:ch2_free_acid_supernatant} to 1D are given in the unitless form as ($\xi_{Cu}\equiv D_{Cu}\tau/H^2\,,$ $\xi_{+}\equiv D_{+}\tau/H^2\,,$ $r\equiv \tau\tilde{r}\,,$ $\xi_{Cu}^{(a)}\equiv D_{Cu}^{(a)}\tau/\alpha^2 H^2\,,$ $\xi_{+}^{(a)}\equiv D_{+}^{(a)}\tau/\alpha^2 H^2$)

\begin{equation}
\begin{split}
     \frac{d\Phi^{(0)}}{d t} &+ \frac{\partial}{\partial Z}\left[\Phi^{(0)}V_f - \xi_{Cu} J^{-1}\partial_z\left(\frac{\Phi^{(0)}}{J}\right)\right] \\& =\frac{1}{J}\left[r\Phi^{(0)}_+\Phi^{(b)}-r\Phi^{(0)}\left(\Phi^*-2\Phi^{(b)}-\Phi^{(b)}_+\right)\right]\,,
\end{split}
\label{eq:ap_free_copper_gel_dl}
\end{equation}

 \begin{equation}
 \begin{split}
      \frac{d\Phi^{(0)}_+}{d t} &+ \frac{\partial}{\partial Z}\left[\Phi^{(0)}_+ V_f - \xi_+ J^{-1}\partial_z\left(\frac{\Phi^{(0)}_+}{J}\right)\right] \\&= -\frac{1}{J}\left[r\Phi^{(0)}_+\left(\Phi^*-\Phi^{(b)}_+\right)\right]\,,
\end{split}
\label{eq:ap_free_acid_gel_dl}
\end{equation}

\begin{equation}
    \frac{d \Phi^{(b)}}{d t} = - \frac{1}{J}\left[r\Phi^{(0)}_+\Phi^{(b)}-r\Phi^{(0)}\left(\Phi^*-2\Phi^{(b)}-\Phi^{(b)}_+\right)\right]\,,
    \label{eq:ap_bound_copper_dl}
\end{equation}

\begin{equation}
    \frac{d \Phi^{(b)}_+}{d t} = \frac{1}{J}\left[r\Phi^{(0)}_+\left(\Phi^*-\Phi^{(b)}_+\right)\right]\,,
    \label{eq:ap_bound_acid_dl}
\end{equation}

\begin{equation}
     \frac{\partial \phi^{(a)}}{\partial t} + \frac{\partial}{\partial z}\left(\phi^{(a)}U - \xi_{Cu}^{(a)}\frac{\partial \phi^{(a)}}{\partial z} \right) = 0\,,
\label{eq:ap_free_copper_supernatant_dl}
\end{equation}

\begin{equation}
     \frac{\partial \phi^{(a)}_+}{\partial t} + \frac{\partial}{\partial z}\left(\phi^{(a)}_+U - \xi_{+}^{(a)}\frac{\partial \phi^{(a)}_+}{\partial z} \right) = 0\,.
\label{eq:ap_free_acid_supernatant_dl}
\end{equation}
    
Eqs.~\ref{eq:ch2_volume_constraint},~\ref{eq:incompressibility_model2}--\ref{eq:ap_free_acid_supernatant_dl} constitute a set of differential algebraic equations that is 10th order in space and seventh order in time. To complete the formulation, the boundary conditions given in Eqs.~\ref{eq:ch2_interface_bc}--\ref{eq:ch2_top_bc} are reduced to 1D as:

\begin{equation}
    \begin{split}
        Z&=0:\quad u_z=0\,,\quad Q_{Cu}=0\,,
        \\&Q_{+} = 
        \begin{cases}
        0\,, & \textit{Scenarios 1, 2\,,}\\ 
        \phi_+^{(f)}U & \textit{Scenario 3\,,}
        \end{cases}
        \\Z&=1:\quad \frac{1}{2}\left(J-J^{-1}\right)=-J\left(\chi\Phi^{(b)}_++\gamma\Phi^{(b)}\right),
        \\&\phi^{(a)} = \frac{\Phi^{(0)}}{J}\,,\quad 
        \phi^{(a)}_+ = \frac{\Phi^{(0)}_+}{J}\,,
        \\&Q_{Cu}^{(a)} = \left(Q_{Cu}+\phi^{(0)}\frac{d u_z}{d t}\right),
        \; Q_{+}^{(a)} = \left(Q_{+}+\phi^{(0)}_+\frac{d u_z}{d t}\right),
        \\z&=1+\alpha:\quad Q_{Cu}^{(a)} =Q_{+}^{(a)}=0\,.
    \end{split}
    \label{eq:ch2_dless_bc}
    \end{equation}
    The seven initial conditions are as follows: In the hydrogel domain, the initial displacement is determined from the experimental observations corresponding to the addition of 1M acid to a hydrogel whose COO$^-$ groups (nominal volume fraction: $\Phi^\ast$) are initially fully occupied with copper ions that can bind to two COO$^-$ groups at once (Sec.~\ref{sec:ap_estimation_para_nl}): 
    \begin{equation}
        u(Z)=-0.0943Z\,, \; \Phi^{(b)} = \frac{\Phi^\ast}{2}\,,\;\Phi^{(b)}_+=\Phi^{(0)}=\Phi^{(0)}_+=0\,.
        \label{eq:ch2_initial_cond}
    \end{equation}
    Eq.~\ref{eq:ch2_initial_cond} gives the initial unitless gel height as $h_0\equiv h(t=0)=1-U(Z=1)=0.9057\,.$ For the addition of the acid to the supernatant domain, the initial free acid and free copper volume fractions are given by ($\phi^{(a)}_{+,i}:$ volume fraction of the acid added to the supernatant) 
    \begin{eqnarray}
        \phi^{(a)}_+(z)&=& 
        \begin{cases}
         \frac{\phi^{(a)}_{+,i}}{2}\left[1+\tanh\left(\Gamma (z-z_0)\right)\right]\,, & \textit{Scenarios 1, 2\,,}\\ 
        0\,, & \textit{Scenario 3\,,}
        \end{cases}   \nonumber
        \\ \phi^{(a)} &=& 0\,,
        \label{eq:ch2_initial_cond2}
    \end{eqnarray}
    The numerical values of the simulation parameters are listed in Table \ref{table:simulation_parameters}. For the nonlinear poroelastic model, we take $\chi=\tilde{\chi}/\bar{p}=3.04$ and $\gamma=\tilde{\gamma}/\bar{p}=6.08\,$ because of the difference in the definition of $\bar{p}$ (Sec. \ref{sec:ap_estimation_para_nl}).

    \subsection{\label{subsec:simulations} Simulation procedure}
    For the numerical implementation of Models I and II, we used the COMSOL 5.4 finite element analysis package~\cite{Comsol}. For Model I, we implemented the weak formulations of the partial differential equations (PDEs) given by Eqs.~\ref{eq:stress_tensor_1D}--\ref{eq:Darcy_flow_1D},~\ref{eq:agent_continuity_model1}~\ref{eq:agent_diffusion_model1}, along with the algebraic equations given by Eqs.~\ref{eq:ch_2_eta_DP}~\ref{eq:ch2_permeability}, subject to the boundary conditions in Eqs.~\ref{eq:pressure_jump_model1},~\ref{eq:BC_model1} and the initial conditions in Eq.~\ref{eq:IC_model1}. For Model II, the weak formulations of the PDEs in Eqs.~\ref{eq:stress_balance_1D},~\ref{eq:Darcy_flow_1D},~\ref{eq:stress_tensor_model2},~\ref{eq:incompressibility_model2},~\ref{eq:ap_free_copper_gel_dl}--\ref{eq:ap_free_acid_supernatant_dl} along with the algebraic equations given by Eqs.~\ref{eq:ch_2_eta_DP}~\ref{eq:ch2_permeability},~\ref{eq:flow}, are subject to the boundary conditions given by the unitless 1D version of the pressure jump condition in Eq.~\ref{eq:ch2_interface_bc} and Eq.~\ref{eq:ch2_dless_bc}, as well as the initial conditions in Eqs.~\ref{eq:ch2_initial_cond},~\ref{eq:ch2_initial_cond2}. To determine $k_{f, min}$ (Eq.~\ref{eq:flow}), we defined a minimum operator in Comsol (minop1) that computes the minimum of $k_f$ over the gel domain as a function of time. For time integration, first- to fifth-order backward differentiation formula was used. The gel and the supernatant solution were assigned separate 1D domains of unit length. The gel domain interval is $Z\in[0, 1]\,,$ whereas the supernatant domain interval $z\in[1+u_z|_{Z=1}/\alpha, 2]$ is approximated to $z\in[1, 2]$ for $u_z\sim \mathcal{O}(1)$ and $\alpha\gg1\,.$ The two domains meet at $Z=z=1$ where continuity conditions are applied.

\section{\label{sec:results} Results}

\subsection{\label{subsec:results_model1} Model I}

\begin{figure}[h]
        \centering
        \includegraphics[width=\columnwidth]{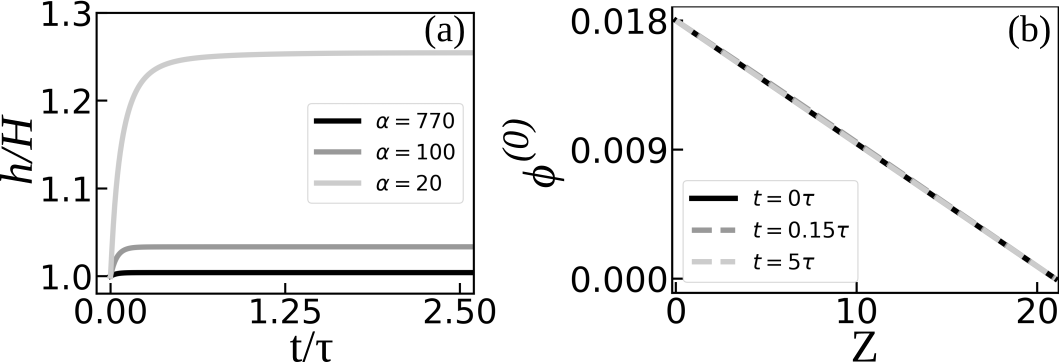}
        \caption{\textbf{Model I: Hydrogel response to imposed agent gradient.} \textbf{(a)} Gel height $h/H\equiv 1+u|_{Z=1}$ versus unitless time for different unitless supernatant domain sizes $\alpha\,,$ with $\alpha=770$ corresponding to the experimental setup in Ref.~\cite{Katke_2024}.  \textbf{(b)} For $\alpha=20\,,$ time evolution of the agent true volume fraction $\phi^{(0)}$ with fixed boundary conditions at $\phi^{(0)}_{in}=0.018$ and $\phi^{(0)}_{out}=0$ (Eq.~\ref{eq:BC_model1}) across the gel and the supernatant domain.}
        \label{fig:model1}
    \end{figure}

The results for Model I that span about two orders of magnitude of the unitless supernatant domain size $\alpha$ are shown in Fig.~\ref{fig:model1}. Because the agent true volume fraction $\phi^{(0)}$ is fixed at the bottom of the gel ($Z=0$) and top end of the supernatant domain ($z=1+\alpha$) as given in Eq.~\ref{eq:BC_model1}, varying $\alpha$ changes the slope of the initial $\phi^{(0)}$ profile (Eq.~\ref{eq:IC_model1}). The imposed agent gradient leads to permanent diffusiophoretic gel deformations where lower $\alpha$ results in both a higher strain and strain rate because of a steeper gradient (Fig.~\ref{fig:model1}a). The maximum limiting strain is achieved when the elastic stress balances the diffusiophoretic stress in Eq.~\ref{eq:gel_displacement_model1} at late times $t/\tau\gg 1\,.$ In this limit, the relative flow speed in the gel $V_f$ vanishes, and Eq.~\ref{eq:agent_continuity_model1} reduces to a diffusion equation in the lab frame. Then, since $\phi^{(0)}$ is merely governed by diffusion per Eqs.~\ref{eq:agent_continuity_model1} and~\ref{eq:agent_diffusion_model1} between two fixed boundary conditions (Eq.~\ref{eq:BC_model1}), it is solved by a linear function across the combined gel and supernatant domain. That is, $\phi^{(0)}$ starts from a linear function (Eq.~\ref{eq:IC_model1}), becomes only slightly disturbed due to the relative influx of the solution in the material frame with lower volume fraction from the supernatant domain that dilutes the solute profile closer to the gel-supernatant interface, and eventually recovers a linear profile because of purely diffusive dynamics (Fig.~\ref{fig:model1}b). Driven by the dilution effect, the convexity for the nominal agent volume fraction $\Phi^{(0)}=J \phi^{(0)}$ becomes pronounced due to the gel swelling that overall increases volumetric factor $J$ in the gel interior (Fig.~\ref{fig:ap_Phi_model1}).

\subsection{\label{subsec:results_model2} Model II}

\begin{figure}[h]
        \centering
        \includegraphics[width=\columnwidth]{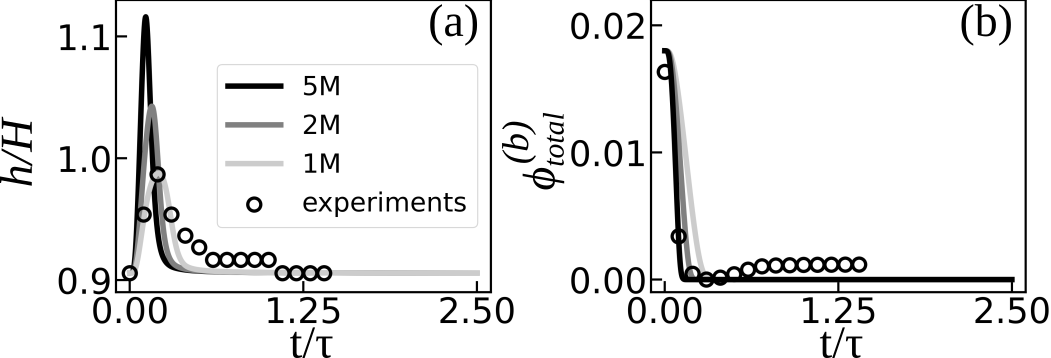}
        \caption{\textbf{Model II, Scenario 1: Hydrogel response to acid.} Hyperelastic model results are shown by the full curves for $\phi_{+,i}^{(a)}=0.006\,, 0.012\,, 0.03\,,$ equivalent to 1M, 2M, and 5M acid, respectively. The dotted data points correspond to the experiments for 1M acid addition, taken from Refs.~\cite{Katke_2024, Korevaar2020}. \textbf{(a)} Gel height $h/H\equiv 1+u|_{Z=1}$ versus unitless time $t/\tau$. \textbf{(b)} Time dependence of the true volume fraction of the total bound agent $\phi^{(b)}_{total}\equiv\int^1_0 \Phi^{(b)}dZ$ for each curve in (a).}
        \label{fig:ch2_figure1}
    \end{figure}

Our results for Scenario 1 are demonstrated in Fig.~\ref{fig:ch2_figure1}. Upon the diffusion of 1M, 2M, and 5M acid, which correspond to $\phi^{(a)}_{+, i}=0.006, 0.012, \text{ and } 0.03,$ respectively, into the agent-laden gel from the supernatant domain, the gel height exhibits a temporal spike with a magnitude $\sim0.1 H$ (Fig.~\ref{fig:ch2_figure1}a). Although the bare rate of rapid gel deformations is given by $\tau_{DP}^{-1}\equiv\tau^{-1}\nu_{DP}\gg \tau^{-1}$ ($\nu_{DP}\sim 10^3$ here; see Table~\ref{table:simulation_parameters}) per Eq.~\ref{eq:gel_displacement_model2}, the effective swelling rate is limited by the overall release time of the total true volume fraction of the bound agent $\Phi^{(b)}_{total}$ (Fig.~\ref{fig:ch2_figure1}b), which is $\tau_{total}/\tau\approx 0.27\,,0.18\,,0.12$ for 1M, 2M, and 5M acid addition, respectively. As a result, the gel undergoes continual diffusiophoretic swelling until $t\gtrsim \tau_{total}$ when the height reaches maximum (Fig.~\ref{fig:ch2_figure1}a). The faster decay of $\phi^{(b)}_{total}$ with increasing acid molarity is due to the larger diffusive influx of acid per Eq.~\ref{eq:ch2_initial_cond2}.  This leads to a steeper gradient of the released agent in the fluid phase with increasing acid molarity and in turn to a higher unitless diffusiophoretic body force defined as $F_{DP}\equiv- \partial(\nu_{DP}\Phi^{(0)})/\partial Z$ (see Eq.~\ref{eq:gel_displacement_model2} and Fig.~\ref{fig:app_figure2}). The increase in $F_{DP}$ improves the average strain rate $s\equiv(h_{max}-h_0)/h_0 t_{max}$ ($h_0:$ initial gel height, $h_{max}:$ height of maximum swelling, $t_{max}:$ time of maximum swelling) up to three times, from $0.44\tau^{-1}$ for 1M acid addition to $2.02\tau^{-1}$ for 5M acid addition (Fig.~\ref{fig:ch2_figure1}). Note that the gel attains its maximum height at a slightly later time $t_{max}$ than the total agent release time $\tau_{max}$ because of the lasting agent diffusion from the hydrogel to the supernatant domain.

\begin{figure}
            \centering
            \includegraphics[width=\columnwidth]{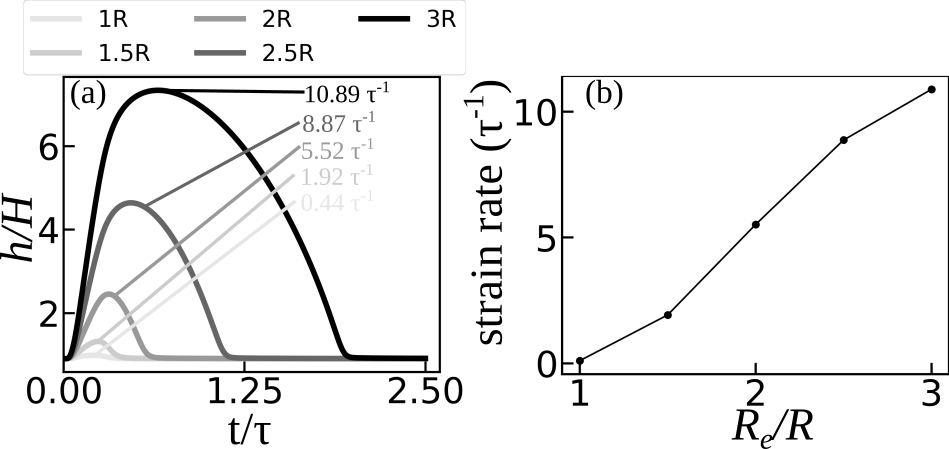}
            \caption{\textbf{Model II, Scenario 2: Gel response to varying polymer-agent interaction strength.} \textbf{(a)} Upon adding 1M acid ($\phi^{(a)}_{+, i}=0.006$), gel height $h/H\equiv 1+u|_{Z=1}$ versus unitless time $t/\tau$ for changing polymer-agent steric repulsion strength when the unitless exclusion radius is varied between $R_e/R=1-3$ where $R=4.24\times 10^{-10}$m is a reference length that corresponds to the exclusion radius between copper divalent ions and PAA polymers (Table~\ref{table:simulation_parameters}). \textbf{(b)} Average strain rates $s\equiv(h_{max}-h_0)/h_0 t_{max}$ extracted from the data in (a) as a function of $R_e/R$ ($h_0:$ initial gel height, $h_{max}:$ height of maximum swelling, $t_{max}:$ time of maximum swelling).}
            \label{fig:ch2_figure5}
        \end{figure}

        We turn to Scenario 2 to study the effect of the unitless diffusiophoretic mobility $\nu_{DP}\equiv k_B T R_e^2/v_c\mu k_f$ on the strain rate by varying the exclusion radius $R_e\,.$ Here we only consider an initial acid concentration at 1M, i.e., $\phi_{+, i}^{(a)}=0.006\,.$ Fig.~\ref{fig:ch2_figure5} demonstrates an increase both in hydrogel swelling strain and the strain rates with bigger $R_e\,.$ whereas the decay of the bound copper volume fraction profiles remains independent of $R_e$ (Fig.~\ref{fig:ap_phi_varying_R}). This behavior can be understood by noting that $\nu_{DP}=R_e^2/2k_f$ in Eq.~\ref{eq:gel_displacement_model2} increases quadratically with the radius of the agent. On the other hand, the agent diffusivity $D$ scales inversely with $R_e$ according to the Stokes-Einstein relation, i.e., $D\sim1/R_e\,.$ Consequently, the released agent volume fraction gradient scales as $\partial \Phi^{(0)}/\partial Z \sim 1/R_e\,.$ Therefore, the unitless diffusiophoretic body force $F_{DP}$ and in turn the average strain rate $s\sim \partial u_z/\partial Z$ per Eq.~\ref{eq:gel_displacement_model2} must both be linear in $R_e$ to leading order, as verified in Fig.~\ref{fig:ch2_figure5}b. This linear scaling enables average strain rates up to $10.89\tau^{-1}\,,$ which is $\sim 25$ times faster than in the linear regime ($R_e/R=1$). 

\begin{figure}
            \centering
            \includegraphics[width=0.65\columnwidth]{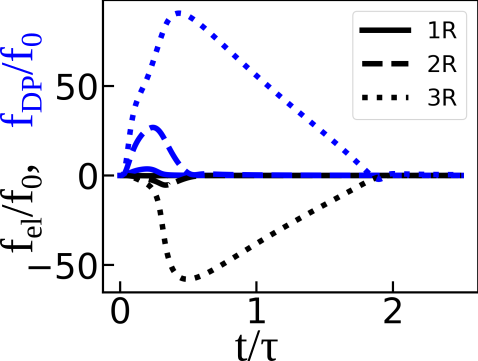}
            \caption{\textbf{Model II, Scenario 2: Diffusiophoretic body force $\boldsymbol{f_{DP}}$ (blue curves) versus elastic body force $\boldsymbol{f_{el}}$ (black curves) for varying polymer-agent interaction strength.} \textbf{(a)} Time evolution of the unitless height-averaged quantities $f_{DP}/f_0$ and $f_{el}/f_0$ corresponding to the gel height profiles in Fig.~\ref{fig:ch2_figure5} for exclusion radii $R_e=R\,, 2R\,, 3R\,.$ Here, $f_0\equiv\bar{p}/H$ is the characteristic body force scale.}
            \label{ch2:diffusive_force_vs_diffusiophoretic_force}
        \end{figure}

     At all molarities and exclusion radii in Figs.~\ref{fig:ch2_figure1} and~\ref{fig:ch2_figure5}, the decay from the maximum height is governed by the competition between the elastic restoration and the residual diffusiophoretic swelling that together lead to a subdiffusive gel relaxation. To understand this competition, we compare the evolution of the height-averaged unitless diffusiophoretic body force $f_{DP}\equiv -\int_{0}^{1} dZ \partial(\nu_{DP}\Phi^{(0)})/\partial Z$ versus the height-averaged unitless elastic body force $f_{el}\equiv \int_{0}^{1} dZ(1/2)\partial\left(J-J^{-1}\right)/\partial Z\,,$ which are nondimensionalized by a characteristic body force scale $f_0\equiv\bar{p}/H$ (see Eq.~\ref{eq:gel_displacement_model2}). In Fig.~\ref{ch2:diffusive_force_vs_diffusiophoretic_force}, $f_{el}$ is predominantly diffusive to the leading order and drives the gel relaxation beyond $t_{max}\,.$ Yet, this elastic restoration is slowed down by the ever damping $f_{DP}>0$, leading to subdiffusive dynamics. This means that superdiffusive swelling is followed by subdiffusive relaxation, pointing to a tradeoff in the swelling-deswelling cycle. 

\begin{figure}
             \centering
             \includegraphics[width=\columnwidth]{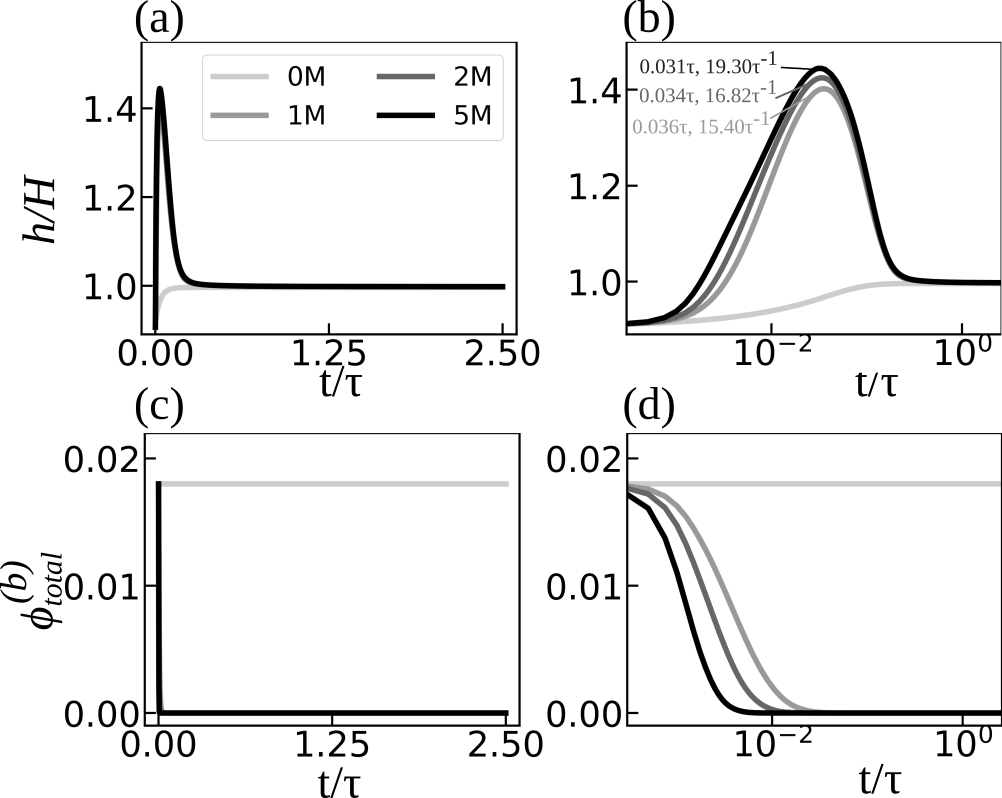}
             \caption{\textbf{Model II, Scenario 3: Gel response to acid flow.} \textbf{(a), (b)} For acid flow into the gel with different volume fractions $\phi_{+}^{(f)}=0.006\,, 0.012\,, 0.03\,,$ (equivalent to 1M, 2M, and 5M acid, respectively), gel height $h/H\equiv 1+u|_{Z=1}$ versus unitless time $t/\tau$. The case for pure solvent flow (i.e., 0M of acid) is also shown for reference. In (b), the maximum swelling time $t_{max}$ and the average strain rate are given for 1M, 2M, and 5M acid. \textbf{(c), (d)} Time dependence of the total bound copper true volume fraction $\phi^{(b)}_{total}=\int^1_0 \Phi^{(b)}dZ$ corresponding to each curve in (a), (b). The semilog plots in panels (b), (d) are the insets of (a) and (c), respectively, and show the gel height at early times in logarithmic scale.}
             \label{fig:ch2_figure3}
     \end{figure}

    To increase the strain rate for small agents ($R_e\sim R$), we consider Scenario 3 that investigates the gel response to acid flux $Q_+=\phi^{(f)}_+U$ into the gel in the positive $Z$ direction, as formulated by Eqs.~\ref{eq:gel_displacement_model2},~\ref{eq:flow}, and~\ref{eq:ch2_dless_bc} (Fig. \ref{fig:ch2_acid_throughflow_schematic}e,~f). We vary the net acid flow rate by varying $\phi^{(f)}_+$ between 0.006 and 0.03 (corresponding to 1M to 5M acid concentrations, respectively)  while keeping the pressure drop in Eqs.~\ref{eq:gel_displacement_model2},~\ref{eq:flow} constant at $\Delta p=0.1$ (Fig.~\ref{fig:ch2_figure3}). The gel exhibits rapid transient swelling, and we record average strain rates up to $19.30 \tau^{-1}$ for 5M acid flow (Fig.~\ref{fig:ch2_figure3}a,~b). This corresponds to about 44 times improvement over Scenario 1. For reference, we also simulate gel dynamics under flow without acid (i.e., 0M concentration), which reaches equilibrium where the flow induced stress is balanced by the gel elasticity and contractile stress by the constant bound agent profile (Fig.~\ref{fig:ch2_figure3}a,~c). Because in our model the contractile stress prefactors $\chi$ and $\gamma$ in Eq.~\ref{eq:stress_tensor_model2} are chosen to give an equal contracted gel height, which is motivated by the experiments in Ref.~\cite{Korevaar2020, Katke_2024}, the diffusiophoretic swelling of the gel under 1--5M acid flow relaxes to the same equilibrium height, this time imposed by the bound-acid induced contractile stresses (Fig.~\ref{fig:ch2_figure3}a).  Note that any hydrogel is essentially a ``porous plug" and can accommodate modest flow speeds. In our case, $\Delta p=0.1$ yields a flow speed $U\approx 0.1U^{(0)}\equiv0.1\mu \text{m/s}$ well below the velocity scale of the gel $U^{(0)}\equiv k_{f,0}\bar{p}/\mu_fH\,,$ rendering our simulations physical. Nonetheless, the Pecl\'{e}t number inside the gel is $Pe\equiv U H/D^{(0)}\sim 10^{-3}\,,$ that is, upon arriving in the gel, the dynamics of acid will always remain diffusive regardless of the flow rate. The advantage of Scenario 3 lies in continuous acid supply during gel deformation that dissolves the bound agent from the gel backbone at elevated rates (Fig.\ref{fig:ch2_figure3}d), leading to a steeper free agent gradient in the gel. This amplifies the strain rate induced by the gel diffusiophoresis.

\section{\label{sec:conclusion} Discussion and conclusions}

Hydrogel diffusiophoresis provides a way to change the rate of hydrogel deformations by overcoming the diffusive scaling without any structural modifications to a gel. Therefore, it offers a fundamentally different physical mechanism than typical engineering approaches, such as increasing pore size (i.e., higher hydraulic permeability $k_f$) or the gel elastic modulus that linearly increase the strain rate within the bounds of diffusive scaling, while compromising on gel functionalization and responsiveness to external fields due to reduced polymer density~\cite{Choudhary2022, epstein2014, Arens2017}. Models I and II based on our hyperelastic theory offer ways to induce hydrogel diffusiophoresis via an external agent gradient (Model I) or internally generated agent gradient (Model II), showcasing the potential versatility of this phenomenon illustrated in Fig.~\ref{fig:diffusiophoresis}.

The theory for gel diffusiophoresis can be generalized to polyelectrolyte hydrogels, where swelling is more pronounced than in regular gels due to the Gibbs-Donnan effect~\cite{Doi2013}. Note that although the PAA hydrogel considered in Model II and in our earlier works~\cite{Korevaar2020, Katke_2024} is a polyelectrolyte gel at neutral or basic pH, here we considered its uncharged state at low pH where the COO$^-$ groups are always neutralized by either the bound copper ions or acid (Fig.~\ref{fig:ch2_acid_throughflow_schematic}b--d). Realizing gel diffusiophoresis in polyelectrolyte gels, however, would reveal an untapped potential for even higher strain rates and power output. The large polyelectrolyte gel deformations are driven by the electrostatic interactions between a charged polymer backbone and a coion and counterion distribution dissolved in the gel. To this end, the theory requires three modifications to the one presented here: First, instead of the linear dependence in Eq.~\ref{eq:enthalpic_stress}, the diffusiophoretic stress should scale logarithmically with the agent volume fraction $\phi^{(0)}$ since the Debye screening length depends on it~\cite{Marbach2019}. Second, there will be an electrostatic potential governed by a Poisson's equation that emerges from the free and bound charge interactions. Third, each of the free coion and counterion distributions will obey a Nernst-Planck equation with an electrostatic flux term. Currently, we are developing a hyperelastic theory for polyelectrolyte gel diffusiophoresis by incorporating these three factors to describe potentially elevated strain rates in this system, to be published elsewhere.

Another case to be explored is collective gel diffusiophoresis in the presence of multiple diffusiophoretic agents. When there is only one diffusiophoretic agent present, superdiffusive swelling in Figs.~\ref{fig:ch2_figure1},~\ref{fig:ch2_figure5},~\ref{fig:ch2_figure3} is always followed by a subdiffusive contraction due to the competition between the elastic and diffusiophoretic body forces (Fig.~\ref{ch2:diffusive_force_vs_diffusiophoretic_force}). Making the relaxation stage superdiffusive would necessitate a second diffusiophoretic agent with an opposite effect on the gel strain, i.e., the second one driving contraction if the first one causes swelling, or vice versa (Fig.~\ref{fig:diffusiophoresis}). Importantly, the activation of the second agent needs to be out of phase with the first one, such as through a chemical feedback between the two agents, to realize a completely superdiffusive swelling-contraction cycle. This would be a prerequisite to engineer superdiffusive periodic actuations faster than those in Belousov–Zhabotinsky gels or pH-responsive gels coupled to pH oscillation reactions~\cite{Murase2009, Balazs2013, epstein2014, Aizenberg2018, Yoshida1995, Bilici2010}.

Even though the fact that a linear chemical gradient $\nabla\phi$ underlies polymer diffusiophoresis is well established~\cite{Hinestrosa2020}, a potential shortcoming of the current theory is that the diffusiophoretic mobility (Eqs.~\ref{eq:enthalpic_stress},~\ref{eq:ch_2_eta_DP}) is based on colloidal diffusiophoresis that assumes scale separation between the sizes of the diffusio-osmotic agent (small ions) and colloidal particles. This approximation is nevertheless plausible for uniaxial gel deformations where the projected polymer length along the $z-$axis is much bigger than the agent size, as evidenced by the agreement between theory and experiments in Fig.~\ref{fig:ch2_figure1} (see also Ref.~\cite{Katke_2024}). Yet, it may break down for more complex gel deformation patterns where the size distribution of the disordered gel structure must affect the diffusiophoretic mobility. In fact, molecular dynamics simulations for short polymer diffusiophoresis found that the mobility depends on the chain length and exhibits a nonmonotonic dependence on the monomer-agent interaction strength~\cite{Hinestrosa2020}. Therefore, mesoscale simulations are needed that can model the entire gel domain while explicitly resolving the gel polymer network and monomer-agent interactions without any a priori assumption of the diffusiophoretic stress (Eq.~\ref{eq:enthalpic_stress}) and mobility (Eq.~\ref{eq:ch_2_eta_DP}), and can reach time scales spanning the entire deformation dynamics. Such a computational framework could also guide new analytical theories for the diffusiophoresis of smaller constituents and disordered polymer networks beyond the canonical formulations used in Eq.~\ref{eq:enthalpic_stress} or those for charged systems~\cite{Marbach2019, Prieve_1984}.

Most importantly, both the theory presented herein and additional studies proposed above need a decisive experimental verification of gel diffusiophoresis that complements our understanding from our previous experiments~\cite{Korevaar2020, Katke_2024}. To this end, the most pressing questions are (i) an experimental investigation of Model I that probes the deformations of a generic hydrogel under an imposed agent gradient, and (ii) whether gel diffusiophoresis can overcome the drastic reduction of the strain rate with increasing gel size imposed by the diffusive poroelastic deformation timescale $\tau\,.$ For (i), we are currently developing a microfluidic system that can accommodate $\sim$10--100 micron scale gel slabs, which will also allow us to test (ii). Additionally, we are running experiments on mm-sized PAA gel droplets (orders of magnitude bigger than the 10-micron-thick PAA gel films used in Refs.~\cite{Korevaar2020, Katke_2024}) in order to measure the scaling of the strain rates with gel size to address (ii). Our experimental results will be published elsewhere.
 
In conclusion, we have developed a theory for nonlinear hydrogel diffusiophoresis arising from excluded volume interactions between imposed or released stimulus and the polymer network. Our analysis shows that by integrating gel diffusiophoresis, we achieve strain rates for swelling bursts associated with nonlinear elasticity and superdiffusive behavior. Our theory needs to be validated by experiments for various gel types, diffusio-osmotic agents, as well as mesoscale simulations without an a priori integration of diffusiophoretic terms into the model. Nevertheless, our findings provide a strong theoretical basis for a versatile proof-of-concept material platform with an inherently compliant fast actuation.

\begin{acknowledgments}
We thank A. Alexeev, N. Alizade, J. Barone, S. Cheng, J. Gray, P. A. Korevaar, M. Pleimling for fruitful discussions and the VT College of Science for financial support. We acknowledge the VT Advanced Research Computing Center for computing resources.
\end{acknowledgments}

\appendix

\renewcommand{\thefigure}{A\arabic{figure}}
\setcounter{figure}{0}

\section{Derivations and estimation of parameters}
\label{ase:app_one_sect_1}

        \begin{figure}
            \centering
            \includegraphics[width=0.7\columnwidth]{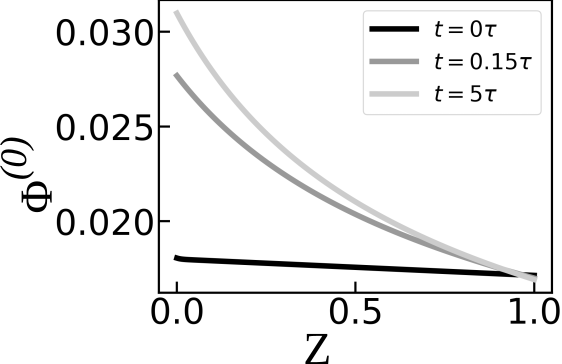}
            \caption{\textbf{Model I: Time evolution of nominal agent volume fraction across the gel.} }
            \label{fig:ap_Phi_model1}
        \end{figure}

            \begin{figure}
             \centering
             \includegraphics[width=\columnwidth]{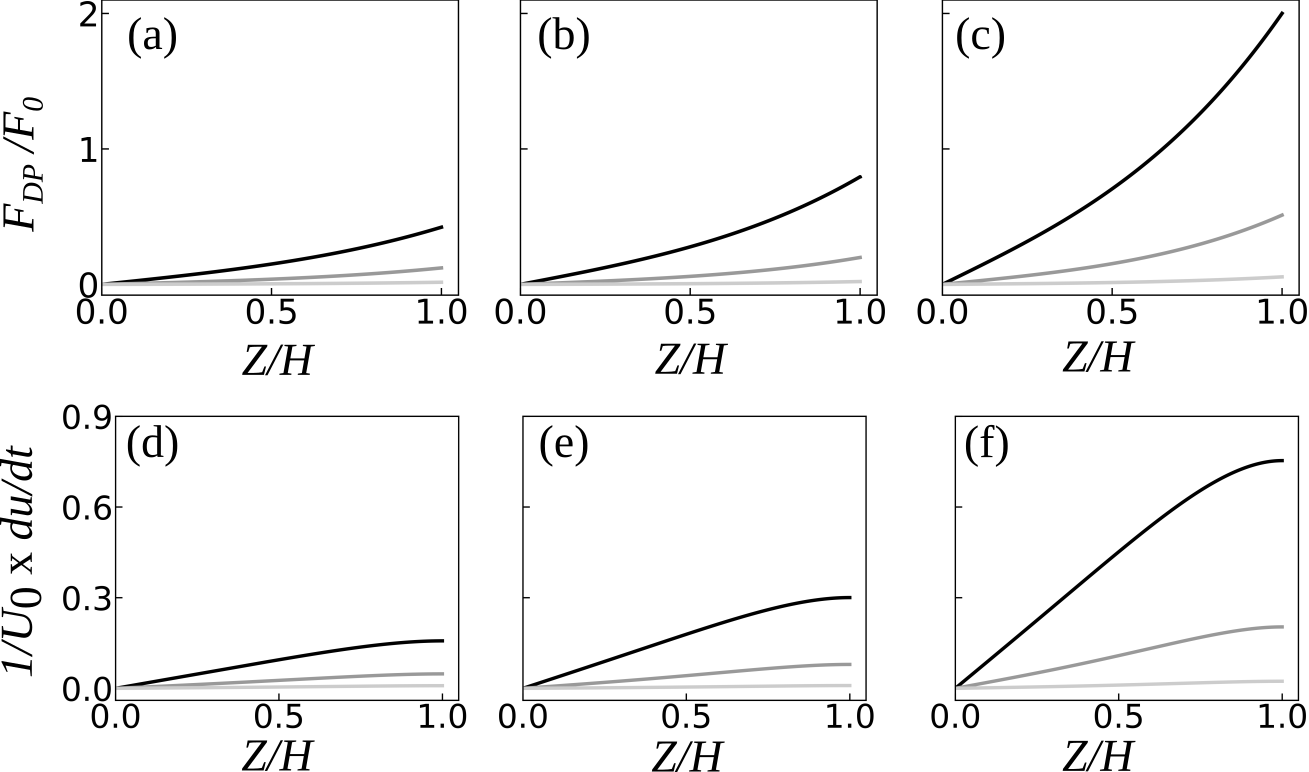}
             \caption{\textbf{Diffusiophoretic body force and strain rate across the gel}. \textbf{(a)}, \textbf{(b)}, and \textbf{(c)} show dimensionless diffusiophoretic body force $F_{DP}/F_0=\frac{k_BTH}{v_c\bar{p}}\frac{\partial(\eta_{DP}\Phi^{(0)})}{\partial Z}$ as function of the Lagrangian coordinate $Z$ for addition of 1M, 2M, and 5M acid respectively at three different times $0.1\,\tau\,,$ $0.15\,\tau\,,$ and $0.25\,\tau\,.$ The corresponding dimensionless strain rate $d\varepsilon/dt$ as function of $Z$ are shown in \textbf{(d)}, \textbf{(e)}, and \textbf{(f)}.}
             \label{fig:app_figure2}
         \end{figure}

        \begin{figure}
            \centering
            \includegraphics[width=0.7\columnwidth]{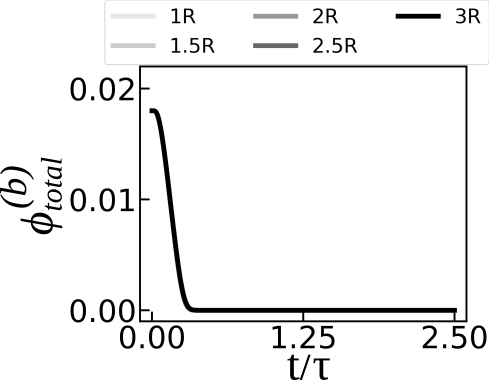}
            \caption{\textbf{Evolution of bound agent volume fraction for different $R_e$.} Upon adding 1M acid ($\phi^{(a)}_{+, i}=0.006$), the true volume fraction of the total bound agent $\phi^{(b)}_{total}\equiv\int^1_0 \Phi^{(b)}dZ$ versus unitless time $t/\tau$ for changing polymer-agent steric repulsion strength when the unitless exclusion radius is varied between $R_e/R=1-3\,.$ Here, $R=4.24\times 10^{-10}$m is a reference length that corresponds to the exclusion radius between copper divalent ions and PAA polymers (Table~\ref{table:simulation_parameters}).}
            \label{fig:ap_phi_varying_R}
        \end{figure}

\subsection{Stress balance in material frame}
    \label{sec:ap_stress_mat_frame}
    
            To derive the stress balance in the material frame, we will use the identity given by Eq.~\ref{eq:stress_conversion} where the true stress (the stress tensor in lab frame) is denoted by $\boldsymbol{\Sigma}$ and the nominal stress (the stress tensor in the material frame) by $\boldsymbol{\sigma}\,.$ Momentum conservation in the lab frame demands 
            \begin{equation}
                \nabla_x \cdot \boldsymbol{\Sigma}=0\,; \quad \frac{\partial \Sigma_{im}}{\partial x_i}=0\,,
                \label{eq:ap_ch4_stress_balance_lab_frame}
            \end{equation}
            where
            \begin{equation}
                    \frac{\partial \Sigma_{im}}{\partial x_i} =\mathbf{F}^{-1}_{li}\frac{\partial}{\partial X_l}\left(\frac{1}{\det \mathbf{F}}\sigma_{mk} F_{ik}\right) = 0\,, \quad \because \Sigma_{im}=\Sigma_{mi}\,.\\
                \label{eq:app_stress_balance}
            \end{equation}
            Expanding Eq.~\ref{eq:app_stress_balance}:
            \begin{equation}
                \begin{split}
                \Rightarrow 0 =&\mathbf{F}^{-1}_{li} F_{ik} \frac{1}{\det \mathbf{F}} \frac{\partial \sigma_{mk}}{\partial X_l} +\mathbf{F}^{-1}_{li} \frac{\partial}{\partial X_l}\left(\frac{1}{\det \mathbf{F}}\right)\sigma_{mk}F_{ik} \\&+  \mathbf{F}^{-1}_{li} \frac{1}{\det \mathbf{F}}\sigma_{mk}\frac{\partial F_{ik}}{\partial X_l}\,,\\
                \Rightarrow 0=&\delta_{lk} \frac{1}{\det \mathbf{F}} \frac{\partial \sigma_{mk}}{\partial X_l} + \delta_{lk} \frac{\partial}{\partial X_l}\left(\frac{1}{\det \mathbf{F}}\right)\sigma_{mk} \\&+  \mathbf{F}^{-1}_{li} \frac{1}{\det \mathbf{F}}\sigma_{mk}\frac{\partial F_{ik}}{\partial X_l}\,,\\
                 \Rightarrow 0=&\frac{1}{\det \mathbf{F}} \frac{\partial \sigma_{mk}}{\partial X_k} + \frac{\partial}{\partial X_k}\left(\frac{1}{\det \mathbf{F}}\right)\sigma_{mk} \\&+   \mathbf{F}^{-1}_{li} \frac{1}{\det \mathbf{F}}\sigma_{mk}\frac{\partial F_{ik}}{\partial X_l}\,.
                 \label{eq:app_stress_der_1}
                \end{split}
            \end{equation}
            Then, using Eq.~\ref{eq:ap_derivative_of_detF_inv} and multiplying both sides with $\det\mathbf{F}$, we arrive at,

            \begin{equation}      
                  \frac{\partial \sigma_{mk}}{\partial X_k}- \mathbf{F}^{-\top}_{\alpha\beta} \frac{\partial  F_{\alpha \beta}}{\partial X_k}\sigma_{mk}+   \mathbf{F}^{-1}_{li} \sigma_{mk}\frac{\partial F_{ik}}{\partial X_l} =0\,.  
            \end{equation}
            Switching the index pairs $\alpha$ and $\beta$ with $i$ and $l$, respectively, and using the definition of the inverse transpose, we obtain,
            \begin{equation}
            \begin{split}
                \frac{\partial \sigma_{mk}}{\partial X_k} -  \mathbf{F}^{-1}_{li} \frac{\partial  F_{il}}{\partial X_k}\sigma_{mk} +   \mathbf{F}^{-1}_{li} \sigma_{mk}\frac{\partial F_{ik}}{\partial X_l} =0\,.
            \end{split}
            \label{eq:app_stress_balance_previous_step}
            \end{equation}
            Using the compatibility of the deformation tensor
            \begin{equation}
                \frac{\partial  F_{ik}}{\partial X_l} = \frac{\partial^2 x_i}{\partial X_k \partial X_l} = \frac{\partial^2 x_i}{\partial X_l \partial X_k} = \frac{\partial  F_{il}}{\partial X_k}
            \end{equation}
            reveals that the second and third terms of Eq.~\ref{eq:app_stress_balance_previous_step} cancel each other out, leading to the momentum conservation in the material frame as
            \begin{equation}          
                \frac{\partial \sigma_{mk}}{\partial X_k}=0\;
                \Rightarrow\; \nabla\cdot\boldsymbol{\sigma}=0\,.     
                \label{eq:app_stress_balance_mat_frame}
            \end{equation}

    \subsection{Fluid and chemical transport in material frame}
    \label{sec:ap_transport_mat_frame}

    While the material frame is convenient to formulate the theory of elasticity for porous media such as hydrogels, fluid and chemical transport is conventionally formulated in the lab frame. Hence, to couple poroelasticity with transport dynamics in the hydrogel, we rewrite the fluid and chemical transport laws in the material frame. 
    
    Let $X_i$ be the position coordinate of the material point in the reference frame defined in a cartesian coordinate system with an orthonormal basis $\{\mathbf{e}_1, \mathbf{e}_2, \mathbf{e}_3\}\,.$ Let $x_i(X_j,t)$ be the current position of the material point also defined in a cartesian coordinate system. Using the Einstein summation convention, the position vectors in the reference frame and the lab frame are given by:
            \begin{equation}
                \mathbf{X} = X_i\mathbf{e}_i\,, \quad \mathbf{x}=x_i(X_j,t)\mathbf{e}_i\,.
            \end{equation}
            We define the displacement vector $\mathbf{u}$ and the deformation tensor $\mathbf{F}$ are as follows:  

            \begin{equation}
                 \mathbf{x} = \mathbf{X} + \mathbf{u}(\mathbf{X,t})\, , \quad  F_{ij} = \frac{dx_i}{dX_j}= \delta_{ij} + \frac{\partial u_i}{\partial X_j}\,. 
            \label{eq:ap_def_tensor2}
            \end{equation}
            Let $\mathbf{dX}$ be the infinitesimal material vector displaced to $\mathbf{dx}$ when acted upon by the deformation tensor $\mathbf{F}\,:$
            \begin{equation}
                \mathbf{dx} = \mathbf{F}\cdot\mathbf{dX}\,, \quad dx_i=F_{ij}dX_j\,.
            \end{equation}
            Let $d\Omega_m$ be the infinitesimal volume in the material frame and $d\Omega$ be the deformed volume in the lab frame. Then,
             \begin{equation}
             \begin{split}
                 d\Omega&=\mathbf{dx_1}\cdot(\mathbf{dx_2}\times \mathbf{dx_3})\\&=\mathbf{F}\cdot\mathbf{dX_1}\cdot(\mathbf{F}\cdot\mathbf{dX_2}\times \mathbf{F}\cdot\mathbf{dX_3})
                 \\&= F_{i\alpha}(dX_1)_\alpha\,\epsilon_{ijk}\,F_{j\beta}(dX_2)_\beta\,F_{k\gamma}(dX_3)_\gamma\\
                &= (\det \mathbf{F})\,(dX_1)_\alpha\,\epsilon_{\alpha\beta\gamma}\,(dX_2)_\beta\,(dX_3)_\gamma\\
                &= (\det \mathbf{F})\,\bigl[\mathbf{dX}_1\cdot(\mathbf{dX}_2\times\mathbf{dX}_3)\bigr]\\
                &= (\det \mathbf{F})\, d\Omega_m = Jd\Omega_m\,. \label{eq:vol_conversion}
            \end{split}
             \end{equation}         
            Here we have used the identity $\epsilon_{ijk}F_{i\alpha}F_{j\beta}F_{k\gamma}=(\det \mathbf{F})\epsilon_{\alpha\beta\gamma}\,.$ We also use a compact notation $J\equiv \det \mathbf{F}\,.$ Using volume conservation between the nominal volume fractions $\Phi$ (defined in the material frame) and true volume fractions  $\phi$ (defined in the lab frame),
            \begin{equation}
            \begin{split}
               \phi d\Omega &= \Phi d\Omega_m\,,\\
               \phi \det \mathbf{F} &= \Phi\,.
               \label{eq:ap_relating_vol_frac_lab_mat}
            \end{split}
            \end{equation}   

           We will give the formulation for reaction and trasport dynamics for Model II, which can straightforwardly be adapted to Model I. Let $\mathbf{v_f}$ be the velocity of the fluid and $\mathbf{v_s}$ be the velocity of the solid phase in lab frame. Let $\phi_s\equiv\phi_p+\phi^{(b)}+\phi^{(b)}_+$ be the volume fraction of the solid phase in the lab frame and $\phi_f=\phi_l + \phi^{(0)} + \phi^{(0)}_+$ be the volume fraction of the fluid phase in the lab frame. To maintain clarity in our derivations, we leave out the rate terms for the conversion of bound copper and bound acid to/from free copper and free acid. Then, the mass conservation equations for the fluid and solid phases in the lab frame are given by

            \begin{equation}
                \frac{\partial \phi_f}{\partial t} + \nabla_x\cdot(\phi_f\mathbf{v_f})=0\,,\quad  \frac{\partial \phi_f}{\partial t} + \frac{\partial (\phi_f\mathbf{v_f})_i}{\partial x_i}=0\,.
                \label{eq:ap_fluid_mass_cons_lab_frame}
            \end{equation}

            \begin{equation}
                \frac{\partial \phi_s}{\partial t} + \nabla_x\cdot(\phi_s\mathbf{v_s})=0\,.
                \label{eq:ap_polymer_cons_lab_frame}
            \end{equation}
            Since $\phi_f + \phi_s=1\,,$ the divergence of the net mass flux in the lab frame vanishes:
            \begin{equation}
                \nabla_x\cdot(\phi_f\mathbf{v_f}+\phi_s\mathbf{v_s})=0\,.
                \label{eq:ap_net_mass_flux_lab_frame}
            \end{equation}
            To derive the fluid transport equation in the material frame, we begin by expanding Eq.~\ref{eq:ap_fluid_mass_cons_lab_frame}.
            \begin{equation}
                \frac{\partial \phi_f}{\partial t} + \mathbf{v_f}_i\frac{\partial \phi_f}{\partial x_i} + \phi_f\frac{\partial \mathbf{v_f}_i}{\partial x_i}=0\,.
            \end{equation}
            We define the relative fluid velocity $\mathbf{v}_i$ as
            \begin{equation}
                \mathbf{v}_i \equiv (\mathbf{v_f}-\mathbf{v_s})_i\,.
            \end{equation}
            Then, the fluid mass conservation equation modifies to
            \begin{equation}
                \underbrace{\frac{\partial \phi_f}{\partial t} + \mathbf{v_s}_i\frac{\partial \phi_f}{\partial x_i}}_{\equiv d^s \phi_f/dt} + \mathbf{v}_i\frac{\partial \phi_f}{\partial x_i} + \phi_f\frac{\partial \mathbf{v}_i}{\partial x_i} + \phi_f\frac{\partial \mathbf{v_s}_i}{\partial x_i} = 0\,.
                \label{eq:app_extra1}
            \end{equation}
            The first two terms on the left-hand side of Eq.~\ref{eq:app_extra1} constitute the time derivative of the fluid phase volume fraction  $\phi_f$ in the material frame. Eq.~\ref{eq:app_extra1} is rewritten as
            \begin{equation}
                    \frac{d^s \phi_f}{dt} +  \frac{\partial (\phi_f\mathbf{v})_i}{\partial x_i} + \phi_f\frac{\partial \mathbf{v_s}_i}{\partial x_i} = 0\,.
                    \label{eq:ap_fluid_flux_der_1}
            \end{equation}
            Next, we define the relative mass flux of the fluid phase in the material frame, $\mathbf{M}$, as follows,
            \begin{equation}
            \begin{split}
                \phi_f \mathbf{v}_i &= \frac{1}{\det \mathbf{F}} F_{il}\mathbf{M}_l\,,\\
                 \mathbf{F}^{-1}_{ni}\det \mathbf{F} \phi_f \mathbf{v}_i &=  F_{ni}^{-1} F_{il}\mathbf{M}_l\,,\\
                 \mathbf{F}^{-1}_{ni} \Phi_f\mathbf{v}_i &= \delta_{nl}\mathbf{M}_l\,,\\
                 \mathbf{F}^{-1}_{ni} \Phi_f\mathbf{v}_i &= \mathbf{M}_n\,,\\
                \Phi_f\mathbf{V_f}_n&=\mathbf{M}_n\,.
                \label{ap:eq_relating_lag_eul_flux}
            \end{split}
            \end{equation}
            Here we define fluid velocity in the material frame as $\mathbf{V_f}_n \equiv  \mathbf{F}^{-1}_{ni}\mathbf{v}_i$ and we have also used the definition of the material fluid volume fraction, $\Phi_f=\phi_f \det \mathbf{F} $ (See Eq.~\ref{eq:ap_relating_vol_frac_lab_mat}). We simplify Eq.~\ref{eq:ap_fluid_flux_der_1} as follows,
            \begin{equation}
                \begin{split}
                    \frac{d^s \phi_f}{dt} +  \frac{\partial (\phi_f\mathbf{v})_i}{\partial x_i} + \phi_f\frac{\partial \mathbf{v_s}_m}{\partial x_m} &= 0\,,\\
                    \frac{d^s \phi_f}{dt} + \underbrace{\frac{\partial X_k}{\partial x_i}\frac{\partial}{\partial X_k}\left(\frac{1}{\det \mathbf{F}} F_{il}\mathbf{M}_l\right)}_{\equiv A_f} + \phi_f\frac{\partial \mathbf{v_s}_m}{\partial x_m} &= 0\,.
                    \label{eq:ap_fluid_flux_der_1.5}
                \end{split}
            \end{equation}
            We expand the second term on the left-hand side of the equation by applying the product rule as follows,
            \begin{equation}
                \begin{split}
                    A_f\equiv&\frac{\partial X_k}{\partial x_i}\frac{\partial}{\partial X_k}\left(\frac{1}{\det \mathbf{F}} F_{il}\mathbf{M}_l\right)\\
                    =& \mathbf{F}^{-1}_{ki}\frac{\partial}{\partial X_k}\left(\frac{1}{\det \mathbf{F}}\right) F_{il}\mathbf{M}_l +  F_{ki}^{-1}\frac{1}{\det \mathbf{F}}\frac{\partial  F_{il}}{\partial X_k} \mathbf{M}_l \\&+  \mathbf{F}^{-1}_{ki}\frac{1}{\det \mathbf{F}} F_{il}\frac{\partial \mathbf{M}_l}{\partial X_k}\,.
                    \label{eq:ap_fluid_flux_der_2}
                \end{split}
            \end{equation}
            Using the identity
            \begin{equation}
                \frac{\partial\det \mathbf{F}}{\partial F_{\alpha \beta}} =  \mathbf{F}^{-\top}_{\alpha\beta} \det \mathbf{F}\,,
                \label{eq:ap_detF_derivative}
            \end{equation}
            the following derivative in Eq.~\ref{eq:ap_fluid_flux_der_2} is rewritten as
            \begin{equation}
                \begin{split}
                    \frac{\partial}{\partial X_k}\left(\frac{1}{\det \mathbf{F}}\right)&=-\frac{1}{(\det \mathbf{F})^2}\frac{\partial}{\partial X_k}(\det \mathbf{F})\\&=-\frac{1}{(\det \mathbf{F})^2}\frac{\partial F_{\alpha\beta}}{\partial X_k}\frac{\partial (\det \mathbf{F})}{\partial  F_{\alpha\beta}} \\& = -\frac{1}{\det \mathbf{F}} \mathbf{F}^{-\top}_{\alpha\beta} \frac{\partial  F_{\alpha \beta}}{\partial X_k}\,.
                    \label{eq:ap_derivative_of_detF_inv}
                \end{split}
            \end{equation}
            Tnen, Eq.~\ref{eq:ap_fluid_flux_der_2} modifies to,
            \begin{equation}
                \begin{split}
                    A_f =&\frac{-1}{\det \mathbf{F}} F_{\alpha\beta}^{-T}\frac{\partial  F_{\alpha\beta}}{\partial X_k} F_{ki}^{-1} F_{il}\mathbf{M}_l +  F_{ki}^{-1}\frac{1}{\det \mathbf{F}}\frac{\partial  F_{il}}{\partial X_k} \mathbf{M}_l \\&+ \frac{1}{\det \mathbf{F}} \mathbf{F}^{-1}_{ki} F_{il}\frac{\partial \mathbf{M}_l}{\partial X_k}\,,
                    \\=&\frac{-1}{\det \mathbf{F}} F_{\alpha\beta}^{-T}\frac{\partial  F_{\alpha\beta}}{\partial X_k}\delta_{kl}\mathbf{M}_l +  F_{ki}^{-1}\frac{1}{\det \mathbf{F}}\frac{\partial  F_{il}}{\partial X_k} \mathbf{M}_l \\&+ \frac{1}{\det \mathbf{F}}\delta_{kl}\frac{\partial \mathbf{M}_l}{\partial X_k}\,,\\
                    =&\frac{-1}{\det \mathbf{F}} F_{\alpha\beta}^{-T}\frac{\partial  F_{\alpha\beta}}{\partial X_k}\mathbf{M}_k +  F_{ki}^{-1}\frac{1}{\det \mathbf{F}}\frac{\partial  F_{il}}{\partial X_k} \mathbf{M}_l \\&+ \frac{1}{\det \mathbf{F}}\frac{\partial \mathbf{M}_k}{\partial X_k}\,.
                \end{split}
            \end{equation}
             Switching the indices $\alpha$ and $\beta$ with $i$ and $l$, respectively, and using the definition of the inverse transpose, we obtain
             \begin{equation}
                 \begin{split}
                    A_f=&-\frac{1}{\det \mathbf{F}} F_{li}^{-1}\frac{\partial  F_{il}}{\partial X_k}\mathbf{M}_k +  F_{ki}^{-1}\frac{1}{\det \mathbf{F}}\frac{\partial  F_{il}}{\partial X_k} \mathbf{M}_l \\&+ \frac{1}{\det \mathbf{F}}\frac{\partial \mathbf{M}_k}{\partial X_k}\,.
                    \label{eq:ap_fluid_flux_der_3}
                 \end{split}
             \end{equation}
            Now we use the compatibility of the deformation tensor, 
            \begin{equation}
                \frac{\partial  F_{ik}}{\partial X_l} = \frac{\partial^2 x_i}{\partial X_k \partial X_l} = \frac{\partial^2 x_i}{\partial X_l \partial X_k} = \frac{\partial  F_{il}}{\partial X_k}\,.
                \label{eq:ap_def_tensor_compatibility}
            \end{equation}
            Hence, Eq.~\ref{eq:ap_fluid_flux_der_3} modifies to,
            \begin{equation}
            \begin{split}
                A_f= &\frac{-1}{\det \mathbf{F}} F_{li}^{-1}\frac{\partial  F_{ik}}{\partial X_l}\mathbf{M}_k + \frac{1}{\det \mathbf{F}}  F_{ki}^{-1}\frac{\partial  F_{il}}{\partial X_k}\mathbf{M}_l \\&+ \frac{1}{\det \mathbf{F}} \frac{\partial \mathbf{M}_k}{\partial X_k}\,.
            \end{split}
            \end{equation}
            Switching the indices $k$ and $l$ in the first term on the right-hand side, we obtain,
             \begin{equation}
                \begin{split}
                   A_f= &\frac{-1}{\det \mathbf{F}} F_{ki}^{-1}\frac{\partial  F_{il}}{\partial X_k}\mathbf{M}_l + \frac{1}{\det \mathbf{F}}  F_{ki}^{-1}\frac{\partial  F_{il}}{\partial X_k}\mathbf{M}_l + \frac{1}{\det \mathbf{F}} \frac{\partial \mathbf{M}_k}{\partial X_k}\\
                   =&\frac{1}{\det \mathbf{F}} \frac{\partial \mathbf{M}_k}{\partial X_k}\,.
                   \label{eq:app_extra2}
                \end{split}    
             \end{equation}
             After substituting $A_f$ (Eq.~\ref{eq:app_extra2}),  Eq.~\ref{eq:ap_fluid_flux_der_1.5} becomes
             \begin{equation}
                \begin{split}
                \frac{d^s \phi_f}{dt} + \frac{1}{\det \mathbf{F}} \frac{\partial \mathbf{M}_k}{\partial X_k} + \phi_f\frac{\partial \mathbf{v_s}_m}{\partial x_m} =& 0\,,\\
                \Rightarrow\frac{d^s \phi_f}{dt} \det \mathbf{F} + \frac{\partial \mathbf{M}_k}{\partial X_k} + \phi_f\frac{\partial \mathbf{v_s}_m}{\partial x_m}\det \mathbf{F} =& 0\,.
                \label{eq:ap_fluid_flux_der_3.5}
                \end{split}
             \end{equation}
             We next show that the last term on the left-hand side of Eq.~\ref{eq:ap_fluid_flux_der_3.5} can be written in terms of the material time derivative of $\det \mathbf{F}$:
             \begin{equation}
                \begin{split}
                 \frac{d^s (\det \mathbf{F})}{dt} = &\frac{d^s(\varepsilon_{ijk} F_{1i} F_{2j} F_{3k})}{dt}\,,\\
                 =&\varepsilon_{ijk}\frac{d^s F_{1i}}{dt} F_{2j} F_{3k} + \varepsilon_{ijk}\frac{d^s F_{2j}}{dt} F_{1i} F_{3k} \\& + \varepsilon_{ijk}\frac{d^s F_{3k}}{dt} F_{1i} F_{2j}\,.
                 \label{eq:ap_fluid_flux_der_4}
                \end{split}
             \end{equation}
             To simplify Eq.~\ref{eq:ap_fluid_flux_der_4}, we use the compatibility condition between the total time derivative and space derivatives in the material frame
             \begin{equation}
             \begin{split}
                 \frac{d^s  F_{1i}}{dt}&=\frac{d^s}{dt}\left(\frac{\partial x_1}{\partial X_i}\right)=\frac{\partial}{\partial X_i}\frac{d^s x_1}{dt} \\&= \frac{\partial \mathbf{v_s}_1}{\partial X_i}=\frac{\partial \mathbf{v_s}_1}{\partial x_m}\frac{\partial x_m}{\partial X_i} = \frac{\partial \mathbf{v_s}_1}{\partial x_m} F_{mi}\,.
             \end{split}
             \end{equation}
             Then, Eq.~\ref{eq:ap_fluid_flux_der_4} simplifies to,
             \begin{equation}
             \begin{split}
                 \frac{d^s (\det\mathbf{F})}{dt} = & \varepsilon_{ijk}\frac{\partial \mathbf{v_s}_1}{\partial x_m} F_{mi} F_{2j} F_{3k} + \varepsilon_{ijk}\frac{\partial \mathbf{v_s}_2}{\partial x_n} F_{nj} F_{1i} F_{3k} \\& + \varepsilon_{ijk} \frac{\partial \mathbf{v_s}_3}{\partial x_p} F_{pk} F_{1i} F_{2j}\,.
                 \label{eq:ap_fluid_flux_der_5}
             \end{split}
             \end{equation}
            Since the determinant of a matrix is zero when two rows are the same, that is, $\varepsilon_{ijk} F_{2i} F_{1j} F_{1k}=\varepsilon_{ijk} F_{2i} F_{2j} F_{3k}=\varepsilon_{ijk} F_{3i} F_{2j} F_{3k}=0$, Eq.~\ref{eq:ap_fluid_flux_der_5} simplifies to,
            \begin{equation}
                \begin{split}
                    \frac{d^s (\det \mathbf{F})}{dt} = & \varepsilon_{ijk}\frac{\partial \mathbf{v_s}_1}{\partial x_1} F_{1i} F_{2j} F_{3k} + \varepsilon_{ijk}\frac{\partial \mathbf{v_s}_2}{\partial x_2} F_{2j} F_{1i} F_{3k} \\& + \varepsilon_{ijk}\frac{\partial \mathbf{v_s}_3}{\partial x_3} F_{3k} F_{1i} F_{2j}\,,\\
                    =& \left(\frac{\partial \mathbf{v_s}_1}{\partial x_1} + \frac{\partial \mathbf{v_s}_2}{\partial x_2} + \frac{\partial \mathbf{v_s}_3}{\partial x_3}\right)\det \mathbf{F}\,,\\
                    =& \frac{\partial \mathbf{v_s}_m}{\partial x_m}\det \mathbf{F}\,.
                \end{split}
            \end{equation}
            Substituting the above relation into Eq.~\ref{eq:ap_fluid_flux_der_3.5} yields
            \begin{equation}
                \begin{split}
                    \frac{d^s \phi_f}{dt} \det \mathbf{F} + \frac{\partial \mathbf{M}_k}{\partial X_k} + \phi_f\frac{\partial \mathbf{v_s}_m}{\partial x_m}\det \mathbf{F} =& 0\,,\\
                    \Rightarrow\frac{d^s \phi_f}{dt} \det \mathbf{F} + \frac{\partial \mathbf{M}_k}{\partial X_k} + \phi_f\frac{d^s(\det \mathbf{F})}{dt} =& 0\,,\\
                    \Rightarrow\frac{d^s (\phi_f \det\mathbf{F})}{dt} +\frac{\partial \mathbf{M}_k}{\partial X_k} =&0\,,\\
                    \Rightarrow\frac{d^s\Phi_f}{dt} + \frac{\partial \mathbf{M}_k}{\partial X_k} =&0\,,\\
                    \Rightarrow\frac{d^s\Phi_f}{dt} + \nabla\cdot(\Phi_f\mathbf{V_f}) =&0\,.
                    \label{ap:eq_net_fluid_mass_cons}
                \end{split}
            \end{equation}

            To describe the reaction and transport dynamics of the fluid phase, we add to Eq.~\ref{ap:eq_net_fluid_mass_cons} the source terms corresponding to the reaction of the free agent ($\Phi^{(0)}$) and free acid ($\Phi^{(0)}_+$) with the polymer backbone: 
            \begin{equation}
                \frac{d^s \Phi_f}{d t} + \nabla\cdot(\Phi_f\mathbf{V_f}) = R_{Cu}-R_+\,.
                \label{ap:eq_net_fluid_mass_cons2}
            \end{equation}
            
            We can transform Darcy's law, which relates the relative fluid flux to the gradient of the pore pressure, i.e.,
            \begin{equation}
                \phi_f\mathbf{v}_i = -\frac{\mathbf{k_f}(\phi_f)}{\mu_f}\cdot\nabla_x p \,,
            \end{equation}
            to the material frame by using Eq.~\ref{eq:ap_def_tensor2}, ~\ref{eq:ap_relating_vol_frac_lab_mat} and~\ref{ap:eq_relating_lag_eul_flux}. After rearranging the terms,  Darcy's law in the material frame becomes
            \begin{equation}
                 \Phi_f\mathbf{V_f} = -\frac{\mathbf{k_f}(\Phi_f/J)}{\mu_f}J\mathbf{F}^{-1}\mathbf{F}^{-\top}\cdot\nabla p\,.
            \end{equation}
            
            We also formulate the transport of the free ions in the material frame. For the free agent transport driven by advection and diffusion, the mass conservation relation in the lab frame is given by,
             \begin{equation}
             \begin{split}
                 \frac{\partial \phi^{(0)}}{\partial t} &+ \nabla_x\cdot(\phi^{(0)}\mathbf{v_f}-D_{Cu}\nabla_x\phi^{(0)})\\&= \underbrace{\tilde{r}\phi^{(0)}_+\phi^{(b)} -\tilde{r}\phi^{(0)}(\phi^*-2\phi^{(b)}-\phi^{(b)}_+)}_{\equiv r_{Cu}}\,.
                 \label{ap:eq_copper_mass_cons_lab_frame}
             \end{split}
             \end{equation}
            In this equation, the source term on the right-hand side is the conversion rate to/from the bound state. We define the relative flux of the agent in the lab frame as,
            \begin{equation}
               \begin{split}
                 \mathbf{q_{r,Cu}} \equiv& \phi^{(0)}\overbrace{(\mathbf{v_f}-\mathbf{v_s})}^{\equiv\mathbf{v}}-D_{Cu}\nabla_x\phi^{(0)}\,.\\
                 \label{ap:eq_relative_copper_flux_lab}
               \end{split}
            \end{equation}
             Then using Eq. \ref{ap:eq_relating_lag_eul_flux}, we can also define the free copper flux in the material frame $\mathbf{\mathbf{Q}_{Cu}}$:
             \begin{equation}
             \begin{split}
                  \mathbf{\mathbf{Q}_{Cu}}_{n} &=  F_{ni}^{-1}\det \mathbf{F} \mathbf{q_{r,Cu}}_i\,,\\
                  &= F_{ni}^{-1}\det \mathbf{F}\left(\phi^{(0)}\mathbf{v}_i-D_{Cu}\frac{\partial \phi^{(0)}}{\partial x_i}\right)\,,\\
                  &= F_{ni}^{-1}\mathbf{v}_i\Phi^{(0)} - D_{cu}  F_{ni}^{-1}\det \mathbf{F} \frac{\partial X_k}{\partial x_i}\frac{\partial \phi^{(0)}}{\partial X_k}\,,\\
                  &= \mathbf{V_f}_n\Phi^{(0)} - D_{Cu}\det \mathbf{F}  \mathbf{F}^{-1}_{ni}  \mathbf{F}^{-1}_{ki} \frac{\partial}{\partial X_k}\left(\frac{\Phi^{(0)}}{\det \mathbf{F}}\right)\,.
            \end{split}
            \end{equation}
            We are now in a position to transform Eq.~\ref{ap:eq_copper_mass_cons_lab_frame} into the material frame:
            \begin{equation}
                \begin{split}
                     \frac{\partial \phi^{(0)}}{\partial t} + \nabla_x\cdot(\phi^{(0)}\mathbf{v_f}-D_{Cu}\nabla_x\phi^{(0)})&= r_{Cu}\,,\\
                     \Rightarrow\frac{\partial \phi^{(0)}}{\partial t} + \mathbf{v_s}_i\frac{\partial \phi^{(0)}}{\partial x_i} + \phi^{(0)}\frac{\partial \mathbf{v_s}_i}{\partial x_i} \\ + \frac{\partial}{\partial x_i}\left(\phi^{(0)}\mathbf{v}_i-D_{Cu}\frac{\partial}{\partial x_i}\phi^{(0)}\right)&= r_{Cu}\,,\\
                     \Rightarrow\frac{d^s \phi^{(0)}}{dt} + \phi^{(0)}\frac{\partial \mathbf{v_s}_i}{\partial x_i} + \frac{\partial \mathbf{m_{r,Cu}}_i}{\partial x_i} &=r_{Cu}\,,\\
                    \Rightarrow\frac{d^s \phi^{(0)}}{dt} + \phi^{(0)}\frac{\partial \mathbf{v_s}_i}{\partial x_i} + \frac{\partial X_k}{\partial x_i} \frac{\partial}{\partial X_k}\left(\frac{F_{il}\mathbf{\mathbf{Q}_{Cu}}_l}{\det \mathbf{F}} \right) &=r_{Cu}\,.
                \end{split}
            \end{equation}
            Following the similar steps as in the derivation of Eq.~\ref{ap:eq_net_fluid_mass_cons}, we arrive at,
            \begin{equation}
            \begin{split}
                 &\frac{d^s \Phi^{(0)}}{d t} + \nabla\cdot\mathbf{Q}_{Cu} = R_{Cu}\,, \quad \text{where}\\
                 &\mathbf{Q}_{Cu} = \Phi^{(0)}_{Cu}\mathbf{V_f} - D_{Cu}J \mathbf{F}^{-1} \mathbf{F}^{-\top}\nabla\left(\frac{\Phi^{(0)}}{J}\right)\,,\\
                 &R_{Cu} = J r_{Cu} = \frac{1}{J}(\tilde{r}\Phi^{(0)}_+\Phi^{(b)}-\tilde{r}\Phi^{(0)}(\Phi^*-2\Phi^{(b)}-\Phi^{(b)}_+))\,.
                 \label{ap:eq_mass_cons_copper_mat}
            \end{split}
            \end{equation}
           Similarly, the mass conservation for free acid in the lab frame is given by: 
           \begin{equation}
                 \frac{\partial \phi^{(0)}_+}{\partial t} + \nabla_x\cdot(\phi^{(0)}_+\mathbf{v_f}-D_{+}\nabla_x\phi^{(0)}_+)= -\tilde{r}\phi^{(0)}_+(\phi^*-\phi^{(b)}_+)\,.
           \end{equation}           
           and in the material frame:
            \begin{equation}
            \begin{split}
                   &\frac{d^s \Phi^{(0)}_+}{d t} + \nabla\cdot\mathbf{Q}_{+} = -R_+\,, \quad \text{where}\\
                   &\mathbf{Q}_{+} = \Phi^{(0)}_{+}\mathbf{V_f} - D_{+}J \mathbf{F}^{-1} \mathbf{F}^{-\top}\nabla\left(\frac{\Phi^{(0)}_+}{J}\right)\,,\\
                   &R_+ = \frac{1}{J}(\tilde{r}\Phi^{(0)}_+(\Phi^*-\Phi^{(b)}_+))\,.
            \end{split}
            \end{equation}

            For bound ion species such as bound agent, mass conservation in the lab frame is given by, 
            \begin{equation}
                \frac{\partial \phi^{(b)}}{\partial t} + \nabla_x\cdot\left(\phi^{(b)}\mathbf{v_s}\right) = -\tilde{r}\phi^{(0)}_+\phi^{(b)} +\tilde{r}\phi^{(0)}\left(\phi^*-2\phi^{(b)}-\phi^{(b)}_+\right)\,.
                \label{ap:eq_bound_ion_transport_lab_frame}
            \end{equation}
           which becomes in the material frame
            \begin{equation}
                \frac{d^s \Phi^{(b)}}{d t} = -R_{Cu}\,.
            \end{equation}
            Similarly, bound acid transport equation in the lab frame is given by, 
            \begin{equation}
                \frac{\partial \phi^{(b)}_+}{\partial t} +  \nabla_x\cdot(\phi^{(b)}_+\mathbf{v}_s)=\tilde{r}\phi^{(0)}_+(\phi^*-\phi^{(b)}_+)\,.
            \end{equation}
            and becomes in the material frame,
            \begin{equation}
                \frac{d^s \Phi^{(b)}_+}{d t} = R_+\,.
            \end{equation}

            Next, we turn to the continuity conditions at the hydrogel-supernatant domain boundary. With $\mathbf{v_s}=d^s \mathbf{u}/d t\,,$ continuity of mass flux at the gel-supernatant boundary in the lab frame can be written as,

            \begin{equation}
                \phi_f\mathbf{v_f} + \phi_s\frac{d^s \mathbf{u}}{d t} = \mathbf{V}\,.
            \end{equation}
            Using Eq.~\ref{ap:eq_relating_lag_eul_flux} and $\phi_f + \phi_s=1\,,$ it can be shown that
            \begin{equation}
                \frac{\mathbf{F}}{J}\cdot(\Phi_f\mathbf{V_f}) + \frac{d^s \mathbf{u}}{d t} = \mathbf{V}\,.
                \label{eq:app_flux_continuity}
            \end{equation}
            
            Flux continuity for free agent at the gel-supernatant boundary in the lab frame is written as,
            \begin{equation}
                (\phi^{(0)}\mathbf{v_f} - D_{Cu}\nabla_x\phi^{(0)})\cdot\mathbf{\hat{n}} = Q^{(a)}\cdot\mathbf{\hat{n}}\,. 
            \end{equation}
            Eqs.~\ref{ap:eq_relative_copper_flux_lab}--\ref{ap:eq_mass_cons_copper_mat} imply that
            \begin{equation}
                \mathbf{Q}_{Cu}^{(a)}\cdot\mathbf{\hat{n}} = \left(\frac{\mathbf{F}}{J}\cdot \mathbf{Q}_{Cu}+\phi^{(0)}\frac{d^s \mathbf{u}}{d t}\right)\cdot\mathbf{\hat{n}}\,,
            \end{equation}
            and for flux continuity of the free acid at gel-supernatant interface,
            \begin{equation}
                \mathbf{Q}_{+}^{(a)}\cdot\mathbf{\hat{n}} = \left(\frac{\mathbf{F}}{J}\cdot \mathbf{Q}_{+} + \phi^{(0)}_+\frac{d^s\mathbf{u}}{dt}\right)\cdot\mathbf{\hat{n}}\,.
            \end{equation}
            We will drop the superscript $s$ from material time derivative to simplify the notation in the main text.                  

            \subsection{Determination of simulation parameters}
            \label{sec:ap_estimation_para_nl}
        We estimate the dimensionless stress coefficients $\chi$ and $\gamma$ for the PAA hydrogel, copper, and acid system~\cite{Korevaar2020, Katke_2024}. When the PAA hydrogel is complexed with copper ions, the height of the hydrogel is stable in the absence of any external stimulus. We use this equilibrium condition to estimate the value of $\gamma\,.$ The stress balance at the gel-supernatant interface implies that
            \begin{equation}
                \frac{1}{2}(J-J^{-1}) = -\gamma\Phi^{(b)}J\,.
                \label{eq:ap_initial_stress_balance_dl}
            \end{equation}
            The initial unitless height of the hydrogel is measured in the experiments to be $h(t=0)=0.9057\,.$ Since this implies $10\%$ strain in the reference equilibrium height, we assume linear deformation, $u(Z,0)=-0.0943Z\,;$ $J=1+\partial u/\partial Z=0.9057$ and uniformity of $\Phi^{(b)}(Z)=0.018$ across the gel. Then Eq. \ref{eq:ap_initial_stress_balance_dl} implies $\gamma=\tilde{\gamma}/\bar{p}=6.08\,.$ Similarly, the gel returns to its initial height after transient swelling as $t \rightarrow \infty$ and $\lim_{t\rightarrow\infty}\Phi^{(b)}_+(Z,t)=0.036\,,$ Hence, the stress balance would imply
            \begin{equation}
                \frac{1}{2}(J-J^{-1}) = -\chi\Phi^{(b)}_+J\,,
            \end{equation}
            and $\chi=3.04\,.$

            To estimate the dimensionless parameters $a$ and $b$ which determine the permeability of the hydrogel (see Eq. \ref{eq:ch2_permeability}), we consider two limits. When $\phi_f\rightarrow1\,,$ $k_f\rightarrow H^2=10^{-10}m^2$ and as $\phi_f\rightarrow 1-\phi_{p, 0}\,,$ the permeability assumes its value in the reference state \cite{Katke_2024}, $k_f\rightarrow k_{f,0}=4\times10^{-19} m^2\,.$ In these limits, $a=1.67\times10^{-10}$ and $b=2.5\times10^8\,.$

        The remaining simulation parameters are listed in Table~\ref{table:simulation_parameters}.

\begin{table*}[t!]
\caption{{\bf Simulation parameters}. The parameters denoted by ${(\ast)}$ correspond to the experimental values in Ref.~\cite{Korevaar2020, Katke_2024}. For the parameters labeled by ${(\dagger)},$ the acid diffusivities are taken to be $D_+=D^{(a)}_+\approx 3\times 10^{-8}m^2/s$ within the experimental range given in Refs.~\cite{Kinght2012_acid1, Agmon1995_acid2}, the copper diffusivities were taken from Ref.~\cite{Wu1990Copper} as $D_{Cu}\approx 6\times 10^{-10}m^2/s$ and $D^{(a)}_{Cu}\approx 10^{-9} m^2/s$ assuming that the Cu$^{2+}$ diffusion in the gel must be slower than in the supernatant due to the polymer presence. Only for Model I, we have used $D_{Cu}=D^{(a)}_{Cu}\approx 10^{-9} m^2/s$ to verify the robustness of a linear agent volume fraction profile (Fig.~\ref{fig:model1}). The diffusiophoretic mobility is estimated as $D_{DP}=k_B T R_e^2/2 v_c\mu_f$ for steric repulsions between the polymer and the Cu$^{2+}$ ions with an exclusion radius $R_e=4.24\times 10^{-10}$~m, equivalent to $\eta_{DP}=R_e^2/2 k_f=0.9$ (Eq.~3). The pressure scale $\bar{p}$ is estimated from Refs.~\cite{Baker2010, Sunyer2012, Denisin2016, Hossain2020}. Among the parameters not labeled by a superscript, $k_f$ and $r$ are chosen in physically plausible ranges, whereas the low aspect ratio $\delta\ll 1$ models the gel as a thin film. The unitless chemical stress moduli $\gamma$ and $\chi$  are found by fitting the experimental gel height of a copper-laden ($t=0$) and fully acid-complexed ($t\rightarrow\infty$) gel film, respectively (Sec.~\ref{sec:ap_estimation_para_nl}).} 
\begin{center}
\small{\begin{tabular}{|c|c|c|}
\hline
\multicolumn{3}{|c|}{{\bf Parameters in real units}}\\   
   \hline
   gel height: $H=10^{-5}$m$~^{(\ast)}$ &  viscosity: $\mu_f=10^{-3}$Pa$\cdot$s$~^{(\ast)}$ & bare permeability: $k_{f, 0}=1\times 10^{-19} m^2$ \\
    & thermal energy: $k_B T\approx 4 \times10^{-21}J$ & molecular volume: $v_c=10^{-29} m^3$$~^{(\ast)}$ \\ 
   \hline 
\end{tabular}}
\vskip 0.1in
\small{\begin{tabular}{|c|c|}
   \hline 
\multicolumn{2}{|c|}{{\bf Poroelastic deformation scales}}\\   
   \hline
   \multicolumn{2}{|c|}{Elastic constants: $\bar{p}\equiv 2\mu=100~\text{kPa}^{(\dagger)},$ } \\ \hline
   Timescale: $\tau\equiv \mu_f H^2/k_f \bar{p}=10 $ s & Diffusivity: $D\equiv H^2/\tau= 1\times 10^{-11} m^2/s$\\ \hline 
\end{tabular}}
\vskip 0.1in
\small{\begin{tabular}{|c|c|c|}   
   \hline
   \multicolumn{3}{|c|}{{\bf Unitless parameters for gel domain}}\\   
   \hline
   H$^{+}$ diffusivity: $\xi_{+}\equiv D_{+}/D=3000$$~^{(\dagger)}$ & reaction rate: $r\equiv \tilde{r}\tau=5\times 10^4$  & mobility: $\nu_{DP}\equiv D_{DP}/D=9\times10^2$$~^{(\dagger)}$\\ 
   Cu$^{2+}$ diffusivity: $\xi_{Cu}\equiv D_{Cu}/D=60$$~^{(\dagger)}$ & &  \\ \hline \multicolumn{2}{|c|}{Ion stress moduli: $\gamma\equiv\tilde{\gamma}/\bar{p}=6.08$ }  & COO$-$ volume fraction: $\phi^\ast=0.036$$~^{(\ast)}$ \\ \multicolumn{2}{|c|}{Acid stress moduli: $\chi\equiv\tilde{\chi}/\bar{p}=3.04$} & polymer volume fraction: $\phi_p=0.04$$~^{(\ast)}$\\\hline
   \multicolumn{3}{|c|}{{\bf Unitless parameters for supernatant domain}}\\ \hline
    \multicolumn{2}{|c|}{Cu$^{2+}$ diffusivity: $\xi^{(a)}_{Cu}\equiv D^{(a)}_{Cu}/\alpha^2 D=1.6\times 10^{-4}$$~^{(\dagger)}$} & height ratio: $\alpha\equiv H^{(a)}/H=770$$~^{(\ast)}$\\ \multicolumn{2}{|c|}{H$^{+}$ diffusivity: $\xi^{(a)}_{+}\equiv D^{(a)}_{+}/\alpha^2 D=5.2\times 10^{-3}$$~^{(\dagger)}$} &\\ \hline
    \end{tabular}}
 \vskip 0.1in
 \small{\begin{tabular}{|c|c|c|}
     \hline
    \multicolumn{3}{|c|}{{\bf Unitless parameters for initial supernatant acid distribution $^{(\ddagger)}$}}\\ \hline
     \multicolumn{3}{|c|}{$\Gamma=1.25\times10^2\,, z_0=1.05\,.$}\\ \hline
    \end{tabular}}
\end{center}
\label{table:simulation_parameters}
\end{table*}


\bibliography{hyperelastic_gel}
\bibliographystyle{unsrt}  

\end{document}